\documentclass[11pt]{article}
\usepackage[a4paper, margin=1in]{geometry}
\usepackage{appendix}

\usepackage[cmex10]{amsmath}
\usepackage{subfig}
\usepackage[numbers]{natbib}
\usepackage{graphicx}
\usepackage{amssymb, amsthm, amsmath}
\usepackage{proof-at-the-end}

\usepackage{mathtools}
\usepackage{commath}
\usepackage{dsfont}
\usepackage{booktabs} 
\usepackage{array} 
\usepackage{paralist} 
\usepackage{verbatim} 
\usepackage{algorithm}
\usepackage{multirow}
\usepackage{algpseudocode}
\let\oldReturn\Return
\renewcommand{\Return}{\State\oldReturn}
\usepackage{diagbox}
\usepackage{tikz}

\newtheorem*{thm*}{Theorem}

\newtheorem*{lem*}{Lemma}

\newtheorem*{pro*}{Proposition}

\newtheorem{definition}{Definition}

\makeatletter
\def\thanks#1{\protected@xdef\@thanks{\@thanks
        \protect\footnotetext{#1}}}
\makeatother

\begin{document}
\title{Revisit 1D Total Variation restoration problem with new real-time algorithms for signal and hyper-parameter estimations}

\author{Zhanhao~Liu$^{1,3,*}$,
        Marion~Perrodin$^{3}$,
        Thomas~Chambrion$^{2}$ 
        and~Radu~S.Stoica$^{1}$
\thanks{* Contact: zhanhao.liu@univ-lorraine.fr}
\thanks{$1$\ 
Université de Lorraine, CNRS, Inria, IECL, F-54000 Nancy, France}
\thanks{$2$\  Institut de Mathématiques de Bourgogne, UMR 5584, CNRS, UBFC, F-21000 Dijon, France.}
\thanks{$3$\ Saint-Gobain Research Paris, 39 Quai Lucien Lefranc, F-93300 Aubervilliers, France.}
}

\maketitle
\begin{abstract}
1D Total Variation (TV) denoising, considering the data fidelity and the Total Variation (TV) regularization, proposes a good restored signal preserving shape edges. The main issue is how to choose the weight $\lambda$ balancing those two terms. In practice, this parameter is selected by assessing a list of candidates (e.g. cross validation), which is inappropriate for the real time application. In this work, we revisit 1D Total Variation restoration algorithm proposed by Tibshirani and Taylor. A heuristic method is integrated for estimating a good choice of $\lambda$ based on the extremums number of restored signal. We propose an offline version of restoration algorithm in $O(n\log n)$ as well as its online implementation in $O(n$). Combining the rapid algorithm and the automatic choice of $\lambda$, we propose a real-time automatic denoising algorithm, providing a large application fields. The simulations show that our proposition of $\lambda$ has a similar performance as the states of the art.  
\end{abstract}

\section{Introduction}
In this article, we consider the denoising of 1D raw signal. Such signals are mathematically modelled with $u$, a real function defined on a bounded open subset $\Omega$ of $\mathbb{R}$: $u:\Omega \rightarrow \mathbb{R}$. Finite number of noised samples of $u$ are available: the noised observations $y = (y_1, \cdots, y_n)$ are sampled at $t = (t_1, \cdots, t_n)$ with $t_1<\cdots<t_n$ and $y_i$ the sample at $t_i$.
We assume the sampling period $\tau_i = t_{i}-t_{i-1}$ is not constant, and the observation $y_i$ is $u(t_i)$ adding an independent random noise $\epsilon$:
\begin{equation}
    y_i = u(t_i) + \epsilon
    \label{equ: obs}
\end{equation}
with $\mathbb{E}(\epsilon) = 0$ and $\mathbb{V}(\epsilon) = \sigma^2$.

\subsection{Signal restoration methods}

Our objective is to restore $u$ by knowing $y$, which is also called \emph{inverse problem}. We hope to have an automatic denoising method in a real time context with high precision and limited calculation resource demand. It is an active research domain, and we give here a short review of existing methods:
\begin{itemize}
    \item Digital filter: for 1D signals with a constant sampling frequency, digital filters are widely used. For example, Savitzky–Golay filter \cite{savitzky1964smoothing} is one of the most popular. It consists in fitting a local polynomial regression to each sample point, and it can be formalized as a convolution with a fixed kernel. The method is extremely rapid, since the kernel can be pre-calculated. However, for the inconstant sampling period, the convolution kernel needs to be recalculated for each sample point, which makes this method inefficient.
    \item Probabilistic approach: by introducing priors regarding the signal properties and the model parameters, the restoration is obtained by maximising the a posteriori distribution. See more details in \cite{Winkler}.
    \item Variational denoising methods: it consists in at first defining an energy function $F(y, u): \mathbb{R}^n \times \mathbb{R}^n \to \mathbb{R}$, and estimating the restored signal by: 
    \begin{equation}
        u^* = \arg \min_u  F(y, u)
    \end{equation}
    Usually, the energy function is a weighted sum of two terms: 
    \begin{equation}
        F(y, u) = L(y,u) + \lambda D(u)
        \label{equ:energie}
    \end{equation}
    with the \emph{data fidelity} $L(y,u)$, the \emph{regularization} $D(u)$ and a hyper-parameter $\lambda\in \mathbb{R}^+$ balancing the weight of two terms.
\end{itemize}

In this article, we focus on variational denoising methods with \emph{Total Variation} (TV) regularization: 
\begin{equation}
    D(u) = \int _{\Omega}|\nabla u|
\end{equation}
where $\nabla u$ is the gradient of $u$.

Total Variation based restoration method is first proposed in \cite{rudin1992nonlinear}. The authors in \cite{chambolle1997image} propose an unconstrained minimisation problem:
\begin{equation}
     u^*_{TV} =  \arg\min_u \int_{\Omega} \lambda|\nabla u| + |u - y|^2
     \label{equ:loss_lions}
\end{equation}
for a given Lagrange multiplier $\lambda>0$, and prove the uniqueness of solution as well as an one-to-one correspondence between the noise's variance $\sigma^2$ and $\lambda$. TV-based methods preserve sharp edges and propose a locally smooth restoration. It has been applied to signal (\cite{harchaoui2010multiple}), image (\cite{rudin1992nonlinear}, \cite{chambolle1997image}) and more generally to graph structure (\cite{ortelli2018total}, \cite{kolmogorov2016total}).

For 1D signal, we can estimate efficiently the exact solution with a given weight parameter $\lambda$. We present here two families:
\begin{itemize}
    \item Dynamic of $\lambda$ (\cite{Tibshirani}, \cite{pollak2005nonlinear}): those algorithms estimate the solution by following the dynamic of $u^*_{TV}$ in function of $\lambda$. The authors in \cite{Tibshirani} propose a \emph{path} algorithm to estimate $\Lambda$, the list of $\lambda$ provoking a change of $u^*_{TV}$ topology, by varying $\lambda$ from $+\infty$ to $0$. Those methods are efficient for estimating the solutions with several candidate values of $\lambda$.
    \item Dynamic of ``new'' sample: for a sequence of $n$ points, the author in \cite{1DVT} proposes to add the samples one by one into consideration and update the solution sequentially based on the local behavior of TV-restoration \cite{louchet:hal-00457763}. This method is particularly appropriate for the online processing of a stream of data.
\end{itemize}

 The performance of the restoration depends on the choice of $\lambda$, which is one of the obstacles for the real application. The automatic hyper-parameters selection is an active research domain for mathematics (probabilistic and variational approaches) and computer science. Traditional approaches consist in calculating a goodness-of-fit criterion for several candidate values and selecting the best one, including cross-validation. Those approaches choose randomly a part of data to test the performance of the model fitted without this part. It is difficult to find out the optimal $\lambda$ in the real time context. Another approach is to introduce some prior knowledge, e.g. in knowing the largest number possible of constant piece of $u$, a good choice of $\lambda$ can be proposed quickly \cite{Tibshirani}. Many efficient methods are based on the prior knowledge about the noise $\epsilon$: the authors in \cite{chambolle2004algorithm} and \cite{wen2011parameter} propose to select the parameter under Morozov's discrepancy principle \cite{morozov2012methods}: e.g. the $L^2$ norm of restoration residuals ($y-u^*_{TV}$) is equal to the variance of noise, and the authors in \cite{hashemi2015adaptive} propose to select the parameter based on the statistical characteristics of restoration residuals. If necessary, we can estimate the noise variance $\sigma^2$ by \cite{donoho1994ideal} and \cite{bioucas2006bayesian}.

We present here more in detail two methods based on known $\sigma^2$ that work well for $n$ points 1D signal:
\begin{itemize}
    \item Stein unbiased risk estimation (SURE) \cite{stein1981estimation} is an unbiased estimator of restoration error $|u^* - u_{net}|^2_2$ with $u_{net}$ the original signal. \cite{Tibshirani} shows that the freedom degree of 1D TV-based restoration is the segment (c.f. Section \ref{sub:model}) number of restored signal, noted $K$, which give us:
	\begin{equation}
		\text{SURE}(\lambda) = |y-u^*(\lambda)|^2_2 + 2\sigma^2 K - n\sigma^2
	\end{equation}

    By assessing a list of candidates, we select the parameter $\lambda$ giving the minimum of $\text{SURE}(\lambda)$:
	\begin{equation}
		\lambda_{\text{SURE}} = \arg \min \text{SURE}(\lambda)
	\end{equation}
	
	\item Adaptive universal threshold (AUT) \cite{sardy2016threshold} selects the parameter $\lambda$ based on the behavior of the restoration with an elegant pre-choice of $\lambda$: $\lambda_N = \sigma/2 \sqrt{n\log\log n}$. Based on the restoration with $\lambda_N$, we get an estimation of the number of segments $\hat K$. Thereafter, the authors adjust the choice of $\lambda$ by:
	\begin{equation}
		\lambda_{AUT} = \frac{\sigma}{2} \sqrt{\frac{n}{\hat{K}}\log\log \frac{n}{\hat{K}}}
	\end{equation}
\end{itemize}

In addition, some heuristic methods are available: the authors in \cite{pollak2005nonlinear} propose to estimate $\lambda$ by tracking the quantity $|u^*_{TV}(\lambda_l) - u^*_{TV}(\lambda_{l-1})|^2$ with $\lambda_l$ the smallest $\lambda$ giving a restoration with $l-1$ segments. More generally for the weight parameter of (\ref{equ:energie}), some heuristic methods (e.g. L-curve \cite{hansen1992analysis}) are proposed based on the variation of $L(y,u)$ in function of $D(u)$.

\subsection{Contributions}
Our contributions are resumed as follows:
\begin{itemize}
    \item First, by revisiting the algorithm proposed in \cite{Tibshirani} and \cite{pollak2005nonlinear}, we propose an online approach to estimate the list $\Lambda$ based on the local influence of a new sample for the TV-based method. This approach is efficient for both hyper-parameter selection and data stream restoration.
    
    \item Second, we propose an automatic choice of the weight parameter $\lambda$ based on the numbers of the restored signal's local extremums in function of $\lambda$ without any prior knowledge on the signal.
\end{itemize}

The overall time complexity (for both the automatic choice of the weight parameter $\lambda$ and the reconstruction of the signal) is in $O(n)$ for online implementation with a space complexity in $O(n)$. 

\subsection{Structure of this paper}

The structure of this work is as follow: in Section \ref{sec:theo}, we will at first revisit the results of \cite{Tibshirani} and \cite{pollak2005nonlinear} about 1D TV-restoration method and analyse the influence of the introduction of a new sample point. Then, based on the established theoretical elements, we propose (Section \ref{sec:algo}) respectively some (offline and online) algorithms for estimating $\Lambda$, $u^*_{TV}$ with a given $\lambda$ and a good choice of hyper-parameter $\lambda$ for a good performance of restoration. Thereafter (Section \ref{sec:res}), we compare our method with different existing methods on some simulated and measured signals. Finally, conclusions and perspectives are depicted. For the sake of readability, the proofs are gathered in Appendix \ref{sec:proofs}.

\section{Theoretical analysis: revisit of 1D Total Variation Restoration}
\label{sec:theo}

We aim to recover the unknown vector $u=(u_1, \cdots, u_n)$ from the noisy observations $y=(y_1, \cdots, y_n)$ with $y_i$ the sample at time $t_i$. We introduce the sampling period vector $\tau = (\tau_1, \cdots, \tau_n)$ with:
\begin{equation}
\tau_i =\begin{cases}
t_{i}-t_{i-1}, & i = 2,..., n.\\
 t_{2}-t_{1}, & i = 1.
\end{cases}
\label{equ:tau}
\end{equation}

We consider the minimization of the discrete approximation of (\ref{equ:loss_lions}):
\begin{align}
    F(\cdot, y, \tau, \lambda):& \mathbb{R}^n \to	 \mathbb{R}\nonumber\\
    & u \mapsto L(y,u,\tau)+\lambda D(u) 
    \label{equ:tv}
\end{align}
with data fidelity term $L(y,u,\tau) = \sum_{i = 1}^n \tau_i(y_i-u_i)^2$ and total variation regularization term $D (u) = \sum_{i=1}^{n-1}|u_i-u_{i-1}|$. For the sake of simplicity, $F(u, y,\tau, \lambda)$ is noted as $F_n(\lambda)$ for the case of $n$ sample points. 

The restored signal is given by: 
\begin{equation}
    u^*(\lambda)=(u^*_1(\lambda), \cdots, u^*_n(\lambda)) = \arg\min F_n(\lambda)
    \label{equ:estimation}
\end{equation}

By following, we will at first introduce the segment notation of the restoration proposed in \cite{pollak2005nonlinear}. Then, we will analyse the dynamic of $u^*(\lambda)$ in function of $\lambda$ and the local modification of the restoration due to the induction of a new sample at the end of sequence.
\subsection{Segment notation}
\label{sub:model}

The considered functional (\ref{equ:tv}) is convex but not derivable. Since we work on a finite sample of signal, the solution $u^*(\lambda)$ can be seen as piece-wise constant. We use a similar notation as \cite{pollak2005nonlinear} to represent the constant piece: a set of index $\{j, j+1, \cdots, k\}$ of consecutive points whose restored value $u^*_j(\lambda) =\cdots = u^*_k(\lambda)$ is called a \emph{segment} if it can not be enlarged, which means if $u^*_{j-1}(\lambda) \neq u^*_j(\lambda)$ (or $j =1$) and $u^*_{k}(\lambda) \neq u^*_{k+1}(\lambda)$ (or $k =n$). The number of segments of $u^*(\lambda)$ is noted as $K(\lambda)$. The following notations are introduced for the $j^{th}$ segment of $u^*(\lambda)$:
\begin{itemize}
    \item Index set $\mathcal{N}_j(\lambda) = \{i^{j}_1, \cdots, i^{j}_{n_j} \}$ with $n_j(\lambda) = \text{Card}(\mathcal{N}_j(\lambda))$, containing the point index inside the $j^{th}$ segment.
    \item Segment level $v^*_j(\lambda) = u^*_i(\lambda), \forall i \in \mathcal{N}_j(\lambda)$ 
\end{itemize}

An equivalent representation of $u^*(\lambda)$ is provided by the \emph{set of segment levels} $v^*(\lambda) =(v^*_1(\lambda), \cdots, v^*_{K(\lambda)}(\lambda))$, with $v^*_1(\lambda) \neq v^*_2(\lambda)$, $v^*_2(\lambda)\neq v^*_3(\lambda)$, $\cdots$, $v^*_{K(\lambda)-1}(\lambda)\neq v^*_{K(\lambda)}(\lambda)$, and the \emph{cutting set} (or signal structure) $\mathcal{N}(\lambda) = \{\mathcal{N}_1(\lambda), \cdots, \mathcal{N}_{K(\lambda)}(\lambda)\}$. The total variation under the segment representation is given by (\ref{equ:vt_v}) which is derivable.
\begin{equation}
    TV(v(\lambda)) = \sum_{j=1}^{K(\lambda)-1} |v_{j+1}(\lambda) - v_j(\lambda)|
    \label{equ:vt_v}
\end{equation}

With a given cutting set $\mathcal{N}(\lambda)$, $v^*(\lambda)$ can be estimated by:
\begin{align}
    v^*(\lambda) &= \{v^*_1(\lambda), \cdots, v^*_{K}(\lambda)\}\nonumber\\ &= \arg\min \{\sum_{j=1}^{K(\lambda)}\sum_{i\in \mathcal{N}_j}\tau_i(y_i-v_j)^2 + \lambda\sum_{j=1}^{K(\lambda)-1} |v_{j+1} - v_j|\}
    \label{equ:solution_v}
\end{align}

Let $s(v) = \{s_0(v), s_1(v), \cdots, s_{K}(v)\}$ with
\begin{equation}
s_i(v) = \begin{cases}
0 &\text{$i=0$ or $K$}\\
\text{sign}(v_{i+1}-v_i) &\text{$i=1,\cdots,K-1$}\\
\end{cases}
\label{equ:sign}
\end{equation}
$K= \text{Card}(v)$ and $s^* = s(v^*(\lambda))$, first order optimality conditions of (\ref{equ:solution_v}) imply:
\begin{equation}
    v^*_j(\lambda) =  \overline{y}^*_j+ \frac{\lambda}{2\mathcal{T}_j} (s^*_{j} - s^*_{j-1})
    \label{equ:v_solution}
\end{equation}
with $j =1,\cdots,K(\lambda)$, the segment length $\mathcal{T}_j = \sum_{i\in \mathcal{N}_j(\lambda)}\tau_i$ and the mean value inside $ j^{th}$ segment $\overline{y}^*_j =  \frac{\sum_{i\in \mathcal{N}_j(\lambda)}\tau_i y_i}{\mathcal{T}_j}$. 

We introduce some items for facilitating the presentation, illustrated in Figure \ref{fig:minmax}:
\begin{itemize}
    \item We will call the local maximum segment \emph{max segment}, the local minimum segment \emph{min segment} and the rest \emph{neutral segment}.
    \item We will call the last point of a segment (excepted the last segment) the \emph{junction} of segments.
\end{itemize}
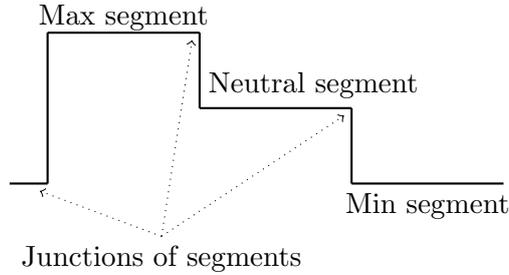
\begin{figure}[!htb]
\centering
\begin{tikzpicture}
\centering
\draw[thick,-] (0,0) -- (0.5,0);
\draw[thick,-] (0.5,0) -- (0.5,2);
\draw[thick,-] (0.5,2) -- (2.5,2);
\draw[thick,-] (2.5,2) -- (2.5,1);
\draw[thick,-] (2.5,1) -- (4.5,1);
\draw[thick,-] (4.5,1) -- (4.5,0);
\draw[thick,-] (4.5,0) -- (6.5,0);
\node at (1.5,2.2){Max segment};
\node at (4,1.3){Neutral segment};
\node at (5.5,-0.3){Min segment};
\draw[dotted,->] (2,-0.7) -- (0.4,-0.1);
\draw[dotted,->] (2,-0.7) -- (2.4,1.9);
\draw[dotted,->] (2,-0.7) -- (4.4,0.9);
\node at (2,-1){Junctions of segments};
\end{tikzpicture}
\caption{Illustration of max, neutral, min segments and the junction of segments. The last point of the last segment is not a junction of segments}
\label{fig:minmax}
\end{figure}

\subsection{Influence of hyper-parameter \texorpdfstring{$\lambda$}{lambda} on cutting set \texorpdfstring{$\mathcal{N}(\lambda)$}{N(lambda)}}

When the cutting set $\mathcal{N}(\lambda)$ is known, the computation of $v^*(\lambda)$ is direct by (\ref{equ:v_solution}). In this section, we present the influence of $\lambda$ on the cutting set $\mathcal{N}(\lambda)$.

The authors in \cite{Tibshirani} and \cite{pollak2005nonlinear} describe the behavior of solution $u^*(\lambda)$ as $\lambda$ decreases. Inspired from an elegant theorem in \cite{kempe2004statistical} and \cite{Winkler} (Proposition 2.5.1, p52) about the piece-wise relation between the restoration with Potts model and the parameter $\lambda$, we reformulate the results in \cite{Tibshirani} by the following theorem: 
\begin{theoremEnd}[no link to proof]{thm}
	There is a sequence $\Lambda = (\lambda_1, \lambda_2, \cdots, \lambda_n, \lambda_{n+1})$ with $0=  \lambda_{n+1} \leq \lambda_n \leq \lambda_{n-1} \leq \cdots \leq \lambda_2< \lambda_1  = \infty$ such that $\forall \lambda_a, \lambda_b \in [\lambda_{l+1}, \lambda_l)$ with $l = 1, \cdots, n$, we have the same cutting set $\mathcal{N}^*(\lambda_a) = \mathcal{N}^*(\lambda_b)$. 
	\label{the:mono_nb_palier}
\end{theoremEnd}

\begin{theoremEnd}[no proof end]{req} 
In the continuous setting, only two segments are merged for $\lambda \in \Lambda$. We have  $0=  \lambda_{n+1} < \lambda_n < \lambda_{n-1} < \cdots < \lambda_2< \lambda_1  = \infty$ such that $\forall \lambda \in [\lambda_{l+1}, \lambda_l)$ with $l = 1, \cdots, n$, the restored signal $u^*(\lambda)$ has $l$ segments: $\text{Card}(\mathcal{N}^*(\lambda))=l$.
\begin{center}
\begin{tikzpicture}
\centering
\draw[thick,->] (0,0) -- (8.5,0);
\draw (0 cm,1pt) -- (0 cm,-1pt) node[anchor=south] {$0$};
\node at (0,-0.5){$\sharp$ segments};
\draw (2 cm,1pt) -- (2 cm,-1pt) node[anchor=south] {$\lambda_n$};
\node at (1.5,-0.5) {$n$};
\draw (3 cm,1pt) -- (3 cm,-1pt) node[anchor=south] {$\lambda_{n-1}$};
\node at (2.5,-0.5) {$n-1$};
\draw (4 cm,1pt) -- (4 cm,-1pt) node[anchor=south] {$\lambda_{n-2}$};
\node at (3.5,-0.5) {$n-2$};
\node at (5,0.25) {$\cdots$};
\draw (5.5 cm,1pt) -- (5.5 cm,-1pt) node[anchor=south] {$\lambda_{4}$};
\node at (6,-0.5) {$3$};
\draw (6.5 cm,1pt) -- (6.5 cm,-1pt) node[anchor=south] {$\lambda_{3}$};
\node at (7,-0.5) {$2$};
\draw (7.5 cm,1pt) -- (7.5 cm,-1pt) node[anchor=south] {$\lambda_{2}$};
\node at (8,-0.5) {$1$};
\end{tikzpicture}
\end{center}
\label{rq:Lambda}
\end{theoremEnd}

\subsection{Dynamic of segments level in function of  \texorpdfstring{$\lambda$}{lambda}}
\label{sub:dynamic}
Theorem \ref{the:mono_nb_palier} allows us to break the estimation of $\Lambda$ into some simpler sub-problems: $\lambda_{l}$ can be estimated from $\lambda_{l+1}$ for $1<l\leq n$. We will describe the dynamic of the restored signal $u^*(\lambda)$ while $\lambda$ increases.

For $\lambda = 0$, (\ref{equ:estimation}) is the least-square estimation, providing $u^*(0) = y$. Assuming that $y_i\neq y_{i+1}$ for every $i$, $u^*(\lambda)$ has $n$ segments for $0 \leq\lambda < \lambda_n$, each segment having only one point.

For simplicity of the presentation, we assume that only two segments are merged for $\lambda\in \Lambda$. Following Remark \ref{rq:Lambda}, $\Lambda$ has $n+1$ distinct elements with $\lambda_{n+1} = 0$ and $\lambda_1 = +\infty$. 

As $\lambda$ increases, (\ref{equ:v_solution}) implies that the min and max segments approach to their neighbors, while the neutral segments remain the same level. For $\lambda \in \Lambda$, two neighbor segments have the same level and form together a new segment. Let's take an example: we assume $\lambda_{l+1}$ known which gives a solution with $l$ segments. Let this solution be $v^l (\lambda)= \{v^l_1(\lambda), \cdots, v^l_l(\lambda)\}$ and $\mathcal{N}^l(\lambda) = \{\mathcal{N}^l_1(\lambda), \cdots, \mathcal{N}^l_l(\lambda)\} $, we introduce also the following notations:
\begin{equation}
\beta_j^l = \frac{1}{2\mathcal{T}^l_j } (s^l_{j} - s^l_{j-1}) 
\label{equ:beta}
\end{equation}
\begin{equation}
\Gamma^l = (\gamma^l_1, \cdots, \gamma^l_{l-1}) = (\beta^l_1 - \beta^l_{2},\cdots, \beta^l_{l-1} - \beta^l_{l})
\label{equ:gamma}
\end{equation}
\begin{equation}
\Delta v^l(\lambda) = (v^l_1(\lambda)-v^l_2(\lambda), \cdots, v^l_{l-1}(\lambda)-v^l_l(\lambda))
\label{equ:delta}
\end{equation}
with $s^l = s(v^l)$ and $ \mathcal{T}^l_j = \sum_{i\in \mathcal{N}^l_j}\tau_i$.

Let $\lambda_{l+1} \leq\lambda_a<\lambda_b<\lambda_{l}$, since the cutting set stays the same, we have: 
\begin{equation}
	 \Delta v^l_i(\lambda_a) = \Delta v^l_i(\lambda_b) +\gamma^l_i(\lambda_a-\lambda_b)
\end{equation}
and $ |\Delta v^l_i(\lambda_a)| \geq |\Delta v^l_i(\lambda_b)|$ for every $i$. The cutting set changes when $\Delta v^l_i(\lambda) = 0$: two segments are merged together. The merged segment index is given by: 
\begin{equation}
    k = \arg\min_k(|\Delta v^l_k(\lambda_{l+1})/ \gamma^l_k|)
    \label{equ:index_merge}
\end{equation}

$\lambda_l$ can be obtained by:
\begin{equation}
    \lambda_l = \lambda_{l+1} + |\Delta v^l_k(\lambda_{l+1})/ \gamma^l_k|
    \label{equ:new_lambda}
\end{equation}
The value of $\lambda_l$ provoking this merge depends on the nature (i.e. min, max or neutral) of those two neighbors ($s^l_{k-1}$, $s^l_{k}$ and $s^l_{k+1}$ ), their length ($\mathcal{T}^l_k$ and $\mathcal{T}^l_{k+1}$) and the difference between their segment levels ($\Delta v^l_k$).
 
The solution with $l-1$ segments, $v^*(\lambda_l)$ and  $\mathcal{N}(\lambda_l)$, is given by the following equations:
\begin{equation}
    v^*(\lambda_l)= \begin{cases}
v^l_j + \beta^l_j(\lambda_{l+1} - \lambda_{l}), &\text{$j=1, \cdots, k-1$}\\
v^l_{j+1} + \beta^l_{j+1}(\lambda_{l+1} - \lambda_{l}), &\text{$j=k,\cdots,l-1$}\\
\end{cases}
    \label{equ:new_v}
\end{equation}
\begin{equation}
    \mathcal{N}(\lambda_l)=\begin{cases}
\mathcal{N}^{l}_j,  &\text{$j=1, \cdots, k-1$}\\
\{\mathcal{N}^{l}_k,\mathcal{N}^{l}_{k+1}\},  &\text{$j=k$}\\
\mathcal{N}^{l}_{j+1}, &\text{$j=k+1,\cdots,l-1$}\\
\end{cases}
    \label{equ:new_n}
\end{equation}

When $\lambda > \lambda_2$, the total variation regularization becomes too important, and all the points form one single segment.

In summary, when $\lambda$ increases, the difference between two neighbour segments levels (i.e. $|v^*_j-v^*_{j+1}|$) becomes smaller, and the total variation of the restored signal decreases.



\subsection{Influence of hyper-parameter \texorpdfstring{$\lambda$}{lambda} on extremums number of restored signal}
 A good choice of $\lambda$ provides a restoration preserving the intrinsic variation of signal and eliminating the local oscillation introduced by noise at the same time.  However, the segment number $\text{Card}(\mathcal{N}^*(\lambda))$ may not be a good indicator of the local oscillation of the restored signal since the neutral segments do not contribute to total variation of restored signal.

Let's return to (\ref{equ:vt_v}): only the max and mix segments are considered in the total variation. We note $g(\lambda)$ the number of extremums of the restored signal $u^*(\lambda)$. A noised signal has a large value of total variation and a large number of local extremums due to the local oscillation, while a well-restored signal which keeps (ideally) only the intrinsic variation of signal may have a small $g(\lambda)$. $g(\lambda)$ seems to us a good indicator of $D(u ^*(\lambda))$ and allows us to estimate a correct choice of $\lambda$ for the restoration.

In this section, we will analyse the relation between $g(\lambda)$ and $\lambda$. Theorem \ref{lem:min_max_palier} shows the monotonicity of the extremums number ($g(\lambda)$) in function of $\lambda$.
\begin{theoremEnd}[no link to proof]{thm}
	There is a sequence $\Lambda^g = (\lambda^g_1, \lambda^g_2, \cdots) \subset \Lambda$ with $\infty=\lambda^g_0  >\lambda^g_1 >\lambda^g_2>\cdots$ such that:
	\begin{itemize}
		\item $g(\lambda_a) = g(\lambda_b)$, $\forall \lambda_a, \lambda_b \in [\lambda^g_{l+1}, \lambda^g_l)$ for every $l$.
		\item $g(\lambda')>g(\lambda'')$, $\forall\lambda' \in [\lambda^g_{l+1}, \lambda^g_l)$ and $\forall\lambda'' \in [\lambda^g_{l}, \lambda^g_{l-1})$ for every $l$.
	\end{itemize}
	\label{lem:min_max_palier}
\end{theoremEnd}
\begin{proofEnd}
  It is easy to show that at least one of the merged segments is a min or max segments since the neutral segment is invariant following $\lambda$. 
  
  Assume that $k+1$ segments are merged into a new segment for a given $\lambda_{l-k} \in \Lambda$. Let the new segment after merging be $\mathcal{N}_j(\lambda_{l-k})=\{\mathcal{N}_{j}(\lambda_l),\cdots, \mathcal{N}_{j+k}(\lambda_l)\}$, we will discuss the following cases:
  
  \begin{itemize} 
  
      \item For $j>1$ and $j+k< K^*(\lambda_l)$, the new segment is neither the first nor the last segment. We discuss the following cases:
      \begin{itemize}
 
      \item $v^*_{j-1}(\lambda_l)>v^*_{j}(\lambda_l)$ and $v^*_{j+k}(\lambda_l)<v^*_{j+k+1}(\lambda_l)$: the new segment is a min segment. It can be easily shown that there must be at least one min segment among $\{\mathcal{N}_j(\lambda_l),\cdots, \mathcal{N}_{j+k}(\lambda_l)\}$, and there must be one more min segments than max segments. So let $m$ the number of min segments among $\{\mathcal{N}_{j}(\lambda_l),\cdots, \mathcal{N}_{j+k}(\lambda_l)\}$, we have $g(\lambda_{l-k}) = g(\lambda_l) - 2(m-1)$.
      
      \item $v^*_{j-1}(\lambda_l)<v^*_{j}(\lambda_l)$ and $v^*_{j+k}(\lambda_l)>v^*_{j+k+1}(\lambda_l)$: the new segment is a max segment. It can be easily shown that there must be at least one max segment among $\{\mathcal{N}_j(\lambda_l),\cdots, \mathcal{N}_{j+k}(\lambda_l)\}$, and there must be one more max segments than min segments. So let $m$ the number of max segments among $\{\mathcal{N}_j(\lambda_l),\cdots, \mathcal{N}_{j+k}(\lambda_l)\}$, we have $g(\lambda_{l-k}) = g(\lambda_l) - 2(m-1)$.
     
      \item $v^*_{j-1}(\lambda_l)> v^*_{j}(\lambda_l)$ and $v^*_{j+k}(\lambda_l)> v^*_{j+k+1}(\lambda_l)$ (or  $v^*_{j-1}(\lambda_l)< v^*_{j}(\lambda_l)$ and $v^*_{j+k}(\lambda_l)< v^*_{j+k+1}(\lambda_l)$): the new segment is a neutral segment. Let $m$ the number of min segments among $\{\mathcal{N}_j(\lambda_l),\cdots, \mathcal{N}_{j+k}(\lambda_l)\}$, we have $g(\lambda_{l-k+1}) = g(\lambda_l) - 2m$.
       \end{itemize}
      \item For $j=1$, the new segment is the first segment, and we discuss the following cases:
      \begin{itemize}
          \item  $v^*_{k+1}(\lambda_l)<v^*_{k+2}(\lambda_l)$: the new segment is a min segment. There must be at least one min segment among $\{\mathcal{N}_1(\lambda_l),\cdots, \mathcal{N}_{k+1}(\lambda_l)\}$, and there may be one more or equal min segments than max segments. So let $m_1$ the number of min segments and $m_2$ the number of max segments among $\{\mathcal{N}_{1}(\lambda_l),\cdots, \mathcal{N}_{k+1}(\lambda_l)\}$, we have $g(\lambda_{l-k}) = g(\lambda_l) - m_1 -m_2 +1$.
          
           \item  $v^*_{k+1}(\lambda_l)>v^*_{k+2}(\lambda_l)$: the new segment is a max segment. There must be at least one min segment among $\{\mathcal{N}_1(\lambda_l),\cdots, \mathcal{N}_{k+1}(\lambda_l)\}$, and there may be one more or equal max segments than min segments. So let $m_1$ the number of min segments and $m_2$ the number of max segments among $\{\mathcal{N}_{1}(\lambda_l),\cdots, \mathcal{N}_{k+1}(\lambda_l)\}$, we have $g(\lambda_{l-k}) = g(\lambda_l) - m_1 -m_2 +1$.
      \end{itemize}
      \item For $j+k= K^*(\lambda_l)$: the new segment becomes the last one. Same as the previous points.
  \end{itemize}
 In summary, $g(\lambda)$ is piece-wise constant and decreasing. 
\end{proofEnd}

$g(\lambda)$ can be estimated directly from $\Lambda$, shown in Proposition \ref{pro:estimation_g}.
\begin{theoremEnd}[no link to proof]{pro}
    Assumes that only two segments are merged together for $\lambda_l \in \Lambda$. In this case, at least one of the two merged segments is a local extremum. Let $\Delta g(\lambda_l) = g(\lambda_{l+1}) -g(\lambda_{l})$:
	\begin{itemize}
		\item If only one segment is an extremum, then the new segment is also an extremum. $\Delta g(\lambda_l) = 0$.
		\item If both segments are extremums and neither are the first or the last segment, the new segment is not an extremum. $\Delta g(\lambda_l) = -2$.
		\item If both segments are extremums and one is the first or the last segment, then the new segment is an extremum. $\Delta g(\lambda_l) = -1$.
	\end{itemize}
	\label{pro:estimation_g}

We propose a simple method to calculate $\Delta g(\lambda_l)$: let $\{v^l, \mathcal{N}^{l}\}$ be the solution for $\lambda = \lambda_{l+1}$, $\mathcal{N}^{l}_j$ and $\mathcal{N}^{l}_{j+1}$ be the two segments to merge, and $s^l = s(v^l)$ following (\ref{equ:sign}), $\Delta g(\lambda_l)$ can be obtained following Table \ref{tab:glambda}.
\begin{table}[!htb]
    \centering
    \begin{tabular}{|c|c||c|}
    \hline
      \multicolumn{2}{|c||}{Conditions}& Results\\
      \hline
          $s^l_{j-1}\times s^l _{j+1}$ & $s^l_{j-1}+ s^l_j+ s^l _{j-1}$ & $\Delta g(\lambda_l)$\\
    \hline
     $\neq0$ & & $=-|s^l_{j-1}+ s^l _{j+1}|$\\\hline
     \multirow{ 2}{*}{$= 0$}&$<2$& $= -1$\\
     \cline{2-3}
     &$=2$& $=0$\\
     \hline
    \end{tabular}
    \caption{Calculation of $\Delta g(\lambda_l)$ in function of  $s^l_{j-1}\times s^l _{j+1}$ and  $s^l_{j-1}+ s^l_j+ s^l _{j-1}$ with $\mathcal{N}^{l}_j$ and $\mathcal{N}^{l}_{j+1}$ the two segments to merge.}
    \label{tab:glambda}
\end{table}
\end{theoremEnd}
\begin{proofEnd}
 Trivial from Theorem \ref{lem:min_max_palier}
\end{proofEnd}

\subsection{Local diffusion of a new observation}
For 1D signal, the new samples arrive sequentially. Assume $n$ samples are collected and the new sample $(y_{n+1}, t_{n+1})$ arrive with $t_{n}<t_{n+1}$. In this section, we show that the restoration change locally with the introduction of the new sample for a fixed $\lambda$.

Let $u^* =(u^*_1, \cdots, u^*_n) = \arg\min F_n$ based on the $n$ first observations with $\lambda$ fixed and $\hat{u}= (\hat{u}_1, \cdots, \hat{u}_n, \hat{u}_{n+1})= \arg\min F_{n+1}$ with $F_{n+1} = F_n + (t_{n+1}-t_n)(y_{n+1}-u_{n+1})^2 + \lambda|u_{n+1}-u_n|$,
we introduce the following notations:
\begin{itemize}
    \item For $\hat{u} = \arg\min F_{n+1}$:
    \begin{itemize}
        \item Segment level set: $\hat{v} = \{\hat{v}_1, \cdots, \hat{v}_{\hat{K}}\}$.
        \item Index set of $j^{th}$ segment: $\hat{\mathcal{N}}_j = \{\hat{i}^{j}_1, \cdots, \hat{i}^{j}_{\hat{n}_j} \}$ with $\hat{n}_j = \text{Card}(\hat{\mathcal{N}}_j)$.
        \item Segment length set: $\hat{\mathcal{T}} = \{\hat{\mathcal{T}}_1, \cdots, \hat{\mathcal{T}}_{\hat{K}}\}$ with $\hat{\mathcal{T}}_j = \sum_{i\in \hat{\mathcal{N}}_j} \tau_i$
    \end{itemize}
    \item For $u^*  = \arg\min F_{n}$:
    \begin{itemize}
        \item Segment level set: $v^* = \{v^*_ 1, \cdots, v^*_{K^*}\}$.
        \item Index set of $j^{th}$ segment: $\mathcal{N}^*_j = \{i^{j, *}_1, \cdots, i^{j, *}_{n^*_j} \}$ with $n^*_j = \text{Card}(\mathcal{N}^*_j)$.
        \item Segment length set: $\mathcal{T}^* = \{\mathcal{T}^*_1, \cdots, \mathcal{T}^*_{K^*}\}$ with $\mathcal{T}^*_j = \sum_{i\in \mathcal{N}^*_j} \tau_i$
    \end{itemize}
\end{itemize}


We can establish the following theorem for the influence of $y_{n+1}$ over the restoration of $n$ first samples: 
\begin{theoremEnd}[no link to proof]{thm}
   
    If there exists an index $j \in \{2, \cdots, K^*(\lambda)\}$ such that $\text{sign}(v^*_{j-1}(\lambda)- v^*_{j}(\lambda))= \text{sign}(v^*_{K^*(\lambda)}(\lambda) - y_{n+1})$, then the new reconstruction $\hat{u}$ satisfies $\hat u_i(\lambda) = u^*_i(\lambda)$ for all $i<i^{j,*}_1$.
    \label{thm:diff_bloque}
\end{theoremEnd}
\begin{proofEnd}
Trivial from Lemma \ref{lem:bloquant}.
\end{proofEnd}

The diffusion from $y_{n+1}$ changes only the last part of $u^*$ up to the junction of two segments whose sign of the variation $v^*_{j-1}(\lambda)- v^*_{j}(\lambda)$ corresponds to that of $v^*_{K^*}(\lambda) - y_{n+1}$, costless to detect. For updating $u^*(\lambda)$ with $y_{n+1}$, only the last points are necessary, and we will introduce the following definition:
\begin{definition}
A sequence $(u^*_m, \cdots, u^*_n)$ is called \emph{non-isolated} with $m = i^{j*}_1$ where $j$ is the last segment which satisfies $\text{sign}(v^*_{j-1}(\lambda)- v^*_{j}(\lambda))= \text{sign}(v^*_{K^*}(\lambda) - y_{n+1})$.
\label{def:non-iso}
\end{definition}

Theorem \ref{thm:len_non_iso} shows the length of the non-isolated sequence decreases in function of $\lambda$. For $\lambda = 0$, each segment contains only one point, and the new sample creates a new segment without any influence to the first $n$ points. For a large value of $\lambda$, all the points form a segment giving the global mean value, so all the restored signal points are influenced by the new sample.
\begin{theoremEnd}[no link to proof]{thm}
	Let $\lambda_1< \lambda_2$ and the length of the non-isolated sequence of $u ^*(\lambda)$ be $l(\lambda)$, we have $l(\lambda_1) \leq l(\lambda_2)$.
	\label{thm:len_non_iso}
\end{theoremEnd}
\begin{proofEnd}
Let $\lambda_1< \lambda_2$, if $u^*_j(\lambda_2) \neq u^*_{j+1}(\lambda_2)$ for $j = 1, \cdots, n-1$, then $u^*_j(\lambda_1) \neq u^*_{j+1}(\lambda_1)$. If the point $j$ is isolated from the influence of new sample for $\lambda = \lambda_2$, then it is also isolated for $\lambda = \lambda_1$.
\end{proofEnd}

In summary, the new sample $(y_{n+1}, t_{n+1})$ changes only the last part of the restoration $u^*(\lambda)$ with a given $\lambda$, and the influence of the new sample diffuses ``further'' with the augmentation of $\lambda$.

\subsection{Independence between segments of restored signal}
\label{sub:inde_lambda}
In this section, we show that the merges of points inside a segment are independent of the points outside the segment.

We propose to save $\Lambda$ in a new vector $\Lambda^{\circ} = (\lambda^{\circ}_1, \lambda^{\circ}_2, \cdots, \lambda^{\circ}_{n-1})$ with $\lambda^{\circ}_i$ the value of $\lambda$ for which the points $i$ and $i+1$ are merged into the same segment. For each segment of the restored signal $u^*(\hat\lambda)$ with a given $\hat \lambda$, we add one point at the junction of segments by following the definition below:


\begin{definition}
  Let $\hat\lambda$  and $\epsilon_\lambda >0$, we have $v^*(\hat\lambda) = \{v_1^*(\hat\lambda), \cdots, v_l^*(\hat\lambda)\}$, $\mathcal{N}^*(\hat\lambda) = \{\mathcal{N}_1, \cdots, \mathcal{N}_l\}$, $l=K(\hat\lambda)$ and $s_i= \text{sign}(v_{i+1}^*(\hat\lambda)-v_i^*(\hat\lambda))$. For each segment $\{y_{\mathcal{N}_j},\tau_{\mathcal{N}_j}\}$ with $j =1, \cdots,l$, let $c_i = \frac{\hat\lambda+\epsilon_\lambda}{2s_i}$, we introduce the \emph{virtual segment} $\{y^+_{\mathcal{N}_j},\tau^+_{\mathcal{N}_j}\}$ where:
  \begin{equation}
  y^+_{\mathcal{N}_i}=\begin{cases}
  \{y_{\mathcal{N}_1}, v_1^*(\hat\lambda) + c_1\}, & i = 1. \\
  \{v_i^*(\hat\lambda) - c_{i-1} , y_{\mathcal{N}_i},  v_i^*(\hat\lambda) + c_i\}, & i = 2,..., l-1.\\
  \{v_{l}^*(\hat\lambda) - c_{l-1}, y_{\mathcal{N}_{l}}\}, & i = l.
  \end{cases}
  \end{equation}
  \begin{equation}
  \tau^+_{\mathcal{N}_i}=\begin{cases}
  \{\tau_{\mathcal{N}_1}, 1\}, & i = 1. \\
  \{1, \tau_{\mathcal{N}_i},  1\}, & i = 2,..., l-1.\\
  \{1, \tau_{\mathcal{N}_{l}}\}, & i = l.
  \end{cases}
  \end{equation}
  
\label{def:virtual}
\end{definition}
Following the dynamic of the restoration described in Section \ref{sub:dynamic}, the variation of segment level depends on the sign of $\Delta v$ but not its value. For a segment of $u^*(\hat \lambda)$ (i.e $j^{th}$ segment $\mathcal{N}^*_j(\hat \lambda) = \{i^{j*}_1, \cdots, i^{(j+1)*}_1-1\}$), the only influences from the points outside the segment are the signs of $\Delta v$ on the borders of the segment (i.e. for the first point of segment $\text{sign}(u^*_{i^{j*}_1}(\hat \lambda) - u^*_{i^{j*}_1-1}(\hat \lambda))$ and for the last point of segment $\text{sign}(u^*_{i^{(j+1)*}_1}(\hat \lambda) - u^*_{i^{(j+1)*}_1-1}(\hat \lambda))$). Under the same border conditions, the merges of sub-segments inside a segment of $u^*(\hat\lambda)$ are independent of the points outside the segment, which will merge with this segment for $\lambda > \hat{\lambda}$. For each virtual segment, those border conditions are guaranteed by the introduction of points at the junction of segments. In consequence, we can break the estimation of $\Lambda^{\circ}$ into some independent sub-problems for each virtual segment, shown in the following proposition:

\begin{theoremEnd}[no link to proof]{pro}
Let $\Lambda^{\circ}$ the estimation with all the samples $\{y_i, t_i\}_{1,\cdots,n}$, $\Lambda^\circ_i= \{\lambda_{i,1}^\circ, \cdots\}$ the estimation with $i^{th}$ virtual segment $(y ^+_{\mathcal{N}_i}, \tau_{\mathcal{N}_i}^+)$ and 
\begin{equation}
\Lambda^{*}_i=\begin{cases}
\{\lambda_{1,1}^\circ, \cdots, \lambda_{1,n_1-1}^\circ\}, & i = 1. \\
\{\lambda_{i,2}^\circ, \cdots, \lambda_{i,n_i}^\circ\}, & i = 2,..., l.\\
\end{cases}
\end{equation}
we have $\cup_{i=1,\cdots, l} \Lambda^{*}_i= \{\lambda |\lambda \leq \hat\lambda \cap \lambda \in \Lambda^{\circ}\}$.
\label{pro:online}
\end{theoremEnd}
\begin{proofEnd}

Let $l$ such that $\lambda_{l+1}\leq \hat \lambda < \lambda_l$. For $\lambda\leq \hat\lambda$, the merge of the elements inside $j^{th}$ segment ($\mathcal{N}^*_{j}(\hat \lambda)$) is independent of those inside $(j-1)^{th}$ ($\mathcal{N}^*_{j-1}(\hat \lambda)$) and $(j+1)^{th}$ segments ($\mathcal{N}^*_{j+1}(\hat \lambda)$) under the same boundary conditions: $\text{sign}(u_{i^j_1}-u_{i^{j-1}_{n_{j-1}}})$ and $\text{sign}(u_{i^{j+1}_1} - u_{i^{j}_{n_{j}}})$ stay the same. The boundary conditions are guaranteed by the introduction of $-(\hat\lambda + \epsilon_\lambda)/(2 s_{i-1}^j)$ and $(\hat\lambda + \epsilon_\lambda)/(2 s_{i}^j)$ since these two new points will merge with their neighbour for $\lambda = \hat\lambda + \epsilon_\lambda$. So we have $\lambda^{\circ, n}_{i^j_1, \cdots, u_{i^{j}_{n_{j}-1}}} = \lambda^{*,j}_{1,\cdots, n_j-1}$.

To finish the proof, notice that $\cup \Lambda^{*,j} = \cup \lambda^{*,j}_{1,\cdots, n_j} = \cup \lambda^{\circ, n}_{i^j_1, \cdots, u_{i^{j}_{n_{j}-1}}} = \{\lambda \leq \hat\lambda, \forall \lambda \in \Lambda^\circ\}$.
\end{proofEnd}

Proposition \ref{pro:online} allows us to estimate separately $\Lambda^\circ$ for each segment of $u^*(\hat{\lambda})$ except the junctions of segments since they have $\lambda^\circ>\hat\lambda$. Combining with the local influence of the new sample to the restoration (i.e $u^*(\hat \lambda)$), the independence between segments implies that the introduction of new sample brings a local modification of $\Lambda^\circ$, respectively the non-isolated sequence of  $u^*(\hat{\lambda})$ and the junction of $u^*(\hat{\lambda})$'s segments. The local influence of the new sample points motivates the online estimation of $\Lambda^\circ$.

\section{Proposition of Algorithms}
\label{sec:algo}
In this section, we will propose some real-time algorithms for estimating a good choice of $\lambda$ and the restoration $u^*(\lambda)$ with a given $\lambda$.

\subsection{Estimation of the elements of \texorpdfstring{$\Lambda$}{Lambda} and their corresponding solution}
\label{sub:DPVT}
The author in \cite{Tibshirani} established a \emph{path} algorithm for estimating $\Lambda$ by moving the parameter from $\lambda = \infty$ to $\lambda = 0$. In the following, we will revisit this algorithm in order to get simultaneously $\Lambda$ and $g(\lambda)$.

Following the dynamic described in Section \ref{sub:dynamic}, the estimation of an element of $\Lambda$ consists in finding two segments to merge by (\ref{equ:index_merge}) and updating all segments in function of $\lambda$ by (\ref{equ:new_v}) and (\ref{equ:new_n}). But the update of the non-merged segments is useless: since $\beta$ does not change, the variation of the segment level $v$ in function of $\lambda$ remains linear with the same slope ($\beta$). 
For estimating $\lambda_l$ giving the solution with $l-1$ segments, the most important is the minimum of $|\frac{\Delta v^l}{\Gamma^l}|$ providing the value of $\lambda_{l}$ and the position of the merge. The merge of two segments changes up to two elements of $\Gamma^l$, respectively the left and right neighbors of the new merged segment. 

Let $\eta = (\eta_1, \cdots, \eta_{n-1})$
with: 
\begin{equation}
    \eta_i =\frac{y_{i+1} - y_{i}}{\beta_{i+1} -\beta_i}  
\end{equation}
and $\beta = (\beta_1, \cdots, \beta_{n})$ where $\beta$ is obtained following (\ref{equ:beta}) with $y$ and $\tau$, we introduce the order of merge $n^\circ = \arg \text{sort} (\eta, \text{ascending = True})$. At each iteration, we need to treat the first element of $n^\circ$, and reproduce $n^\circ$ for the remaining segments after changing the value of $\eta$ for the new segment's neighbor(s).

Besides, at each merge of segments, we save two new values ($\lambda_{new}, \Delta g(\lambda_{new})$), respectively $\lambda_{new}$ the $\lambda$ value provoking the merge and $\Delta g(\lambda_{new})$ the variation of $g(\lambda)$ after the merge. We propose to save those new values in two vectors of size $n-1$ by following Section \ref{sub:inde_lambda}:
\begin{itemize}
	\item $\Lambda^{\circ} = (\lambda^{\circ}_1, \lambda^{\circ}_2, \cdots, \lambda^{\circ}_{n-1})$ with $\lambda^{\circ}_i$ the value of $\lambda$ for which the points $i$ and $i+1$ are merged into the same segment.
	\item $\Delta g^{\circ} = (\Delta g(\lambda^{\circ}_1), \Delta g(\lambda^{\circ}_2), \cdots, \Delta g(\lambda^{\circ}_{n-1}))$ 
\end{itemize}

The algorithm (DP-TV) is shown in Algorithm \ref{algo:Lambda_2}. At each iteration, after finding two merged segments by the first element of $n^\circ$, the new values are saved at the junction of two merged segments (the last point of the left merged segment).

\begin{algorithm}
\caption{DP-TV: Estimation of all elements of $\Lambda$ and their corresponding solution}
\begin{algorithmic} 
\Require $y = (y_1, \cdots, y_n),\ \tau = (\tau_1,\cdots, \tau_n)$

\State $\tilde{v} = y$, $\tilde{\tau} = \tau$
\State $\tilde{\lambda} = (0,\cdots,0)$ \Comment{Initialisation}
\State $s = s(\tilde{v})$ following (\ref{equ:sign})
\State $\tilde{\beta} = \beta$ following (\ref{equ:beta}) with $\tilde{\tau}$ and $s$
\State Left neighbour vector $nl = (0,\cdots, n-1)$
\State Right neighbour vector $nr = (2, \cdots, n+1)$
\State Get $\tilde \Gamma$ and $\Delta \tilde v$ by (\ref{equ:gamma}) and (\ref{equ:delta}) with $\beta$ and $y$
\State $\eta = |\Delta \tilde{v}/ \tilde{\Gamma}|$

\State $n^\circ = \arg \text{sort} (\eta, \text{ascending = True})$

\For{$i = 1, \cdots, n-1$}
\State $k = n^\circ_i$ \Comment{Find segments to merge}
\State $v_{new} = \tilde{v}_k + \tilde{\beta}_k (\eta_k - \tilde\lambda_k)$
\State $ \lambda_{new} = \eta_k$
\State $\tau_{new} = \tilde \tau_k + \tilde \tau_{nr_k}$

\State $\Lambda^\circ_k = \lambda_{new}$ \Comment{Save for output}
\State Get $\Delta g^\circ_k$ by Proposition \ref{pro:estimation_g}
\State $k_{left}, k_{right} = nl_k, nr_k$ \Comment{Update $\eta$ for next merge}
\State $\tilde \lambda_{k_{right}}, \tilde v_{k_{right}}, \tilde \tau_{k_{right}} = \lambda_{new}, v_{new}, \tau_{new}$ 
\State $\tilde \beta_{k_{right}} = \frac{1}{2\tau_{new}}(s_{k_{right}} - s_{k_{left}})$

\State $nl_{k_{right}}, nr_{k_{left}} = k_{left}, k_{right}$
\State $k_{rr} = nr_{k_{right}}$

\If {$k_{left}>=1$} \Comment{Not first segment}
\State $v_{left} = \tilde v_{k_{left}} + (\lambda_{new}-\tilde \lambda_{k_{left}})\tilde \beta_{k_{left}}$
\State $\eta_{k_{left}}=|\frac{ v_{left}- v_{new}}{\tilde \beta _{k_{left}} - \tilde \beta_{k_{right}}}| + \lambda_{new}$
\EndIf
\If {$k_{right}<n$} \Comment{Not last segment}

\State $v_{rr} = \tilde v_{k_{rr}} + (\lambda_{new}-\tilde \lambda_{k_{rr}})\tilde \beta_{k_{rr}}$
\State $\eta_{k_{right}}=|\frac{v_{new}-v_{rr} }{\tilde \beta _{k_{right}} - \tilde \beta_{k_{rr}}}| + \lambda_{new}$
\EndIf
\State Resort $n^\circ_{\{i+1, \cdots, n-1\}}$ following $\eta_{{n^\circ}_{\{i+1, \cdots, n-1\}}}$
 \EndFor
 \Return $\Lambda^{\circ}, \Delta g^\circ$

\end{algorithmic}
\label{algo:Lambda_2}
\end{algorithm}
\begin{theoremEnd}[no proof end]{req} 
\begin{enumerate}
    \item With a careful implementation of the computation of $n^\circ$ (i.e. binary search tree), it can be implemented in low time and space complexity, respectively $O(n\log n)$ and $O(n)$. 
    
    \item If the original noised signal has constant pieces (i.e. $\exists i, y_i = y_{i+1}$), we need to replace the constant piece $\{y, \tau\}_{j,\cdots, k}$ by $ \{y, \sum_{i=j}^k \tau_i\}$ before applying our algorithm. Under the assumption of continuous noise, the probability to have a constant piece is equal to 0, but it remains a practical issue of implementation in the real data with actual floating point arithmetic.
    \item A major practical issue is the merge of multiple segments for the same value of $\lambda$ due to the errors inherent to floating point arithmetic of the micro-processor. With a slight modification, DP-TV can deal with this situation, in which the first elements of $\eta$ are equal.
\end{enumerate}
\end{theoremEnd}

\subsection{Estimation of \texorpdfstring{$u^*(\lambda)$}{solution} and \texorpdfstring{$g(\lambda)$}{g l}} 
With $\Lambda^{\circ}$ estimated by Algorithm \ref{algo:Lambda_2}, we need to apply the following method for estimating the solution $u^*(\lambda)$ with a given $\lambda$:
\begin{itemize}
    \item Find the cutting set: $j = \{j_1, \cdots, j_{l+1}\}$ with $j_1 = 0$, 
    $j_{l+1} = n$ and $\lambda_{j_i}^\circ > \lambda$ for $1<i\leq l$.
     \item Initialisation: $v_{output}$ and $\mathcal{T}_{output}$ two vectors of size $l$. For each segment $[j_i+1, j_{i+1}]$ with $1\leq i\leq l$:
    $\mathcal{T}_{output, i} = \sum_{m={j_i+1}}^{j_{i+1}} \tau_m$ and $v_{output, i} = \frac{1}{\mathcal{T}_{output, i}}\sum_{m={j_i+1}}^{j_{i+1}} \tau_m y_m$.
 
    \item Adjust the segment levels with the given $\lambda$: the level of $i^{th}$ segment is given by:
    \begin{equation}
	 v^*_i(\lambda) = v_{output, i} +\beta_i\lambda
    \end{equation}
    with $\beta_i = \frac{1}{2\mathcal{T}_{output,i} } (s_{i} - s_{i-1})$ and $s = s(v_{output})$.
\end{itemize}
The complexity for estimating $u^*(\lambda)$ with a given $\lambda$ is in $O(n)$.

The estimation of $g(\lambda)$ is given by:
\begin{equation}
    g(\lambda) = 1-\sum_{i=1}^{n-1} \Delta g^\circ_i \mathbf{1}_{\{\lambda^\circ_i - \lambda\}}
\end{equation}
with: 
\begin{equation}
\mathbf{1}_\alpha=\begin{cases}
1, & \alpha > 0\\
0 & \alpha \leq 0\\
\end{cases}
\end{equation}

The complexity for estimating $g(\lambda)$ is in $O(n\log n)$ since we need to sort $\Lambda^\circ$. We can reduce it to $O(n)$ by saving directly $\Lambda$ and $\Delta g(\lambda)$ for $\lambda \in \Lambda$ at each iteration, instead of $\Delta g^\circ$.

\subsection{Estimation of \texorpdfstring{$\lambda$}{lambda} from \texorpdfstring{$g(\lambda)$}{g(lambda)}}
\label{sub;g_lambda}
What is a good $\lambda$? Our objective is to denoise the observations $y$. By knowing the true signal $u_{net}$, the mean squared error of the reconstruction $u^*(\lambda)$ is defined by:
\begin{equation}
    \mathcal{R}(\lambda) = \frac{1}{n}||u_{net} - u^*(\lambda)||_2^2
    \label{equ:mse}
\end{equation}
A good $\lambda$ must have a small value of $\mathcal{R}(\lambda)$. The optimal $\lambda$ can be found out by:
\begin{equation}
    \lambda_{op} = \arg\min \mathcal{R}(\lambda)
    \label{equ:l_op}
\end{equation}

Usually, only some prior knowledge or even nothing about $u_{net}$ is known. Here, we propose a rapid deterministic approach to find out a similar value of $\lambda_{op}$ without any prior knowledge about the original signal $u_{net}$ and the noise $\epsilon$. The only assumption we made is following: the noise $\epsilon$ represents the high frequency of the observed signal $y$, while the original signal $u_{net}$ represents the low frequency. This assumption is natural for the signal processing community. 

With a known cutting set, the denoising is simply done by an average over the points inside each segment. A good denoising cutting must regroup enough points in each segment. The value of $\lambda$ need to be chosen carefully: 
\begin{itemize}
    \item A small $\lambda$ provides a fine cutting: averaging locally over a small portion of points limits the denoising effect.
    \item A large $\lambda$ gives a coarse cutting: the structure of $u$ is vanished by the global average over a big portion of points, and it provides a poor denoising performance.
\end{itemize}

The question is how to get a good compromise between local and global behaviors. The key feature is the number of min-max segments $g(\lambda)$ of $u^*(\lambda)$. Following (\ref{equ:v_solution}), the segment level $v$ is piece-wise linear in function of $\lambda$, and $\frac{\partial v^*_j}{\partial \lambda}$ depends on segment length $\mathcal{T}_j$: short segment is more sensitive to the variation of $\lambda$. Supposing a moderate noise is added to the original signal, we can have the following observations:
\begin{itemize}
    \item For a small $\lambda$, the restored signal has a huge number of small min-max segments introduced by noise. A small augmentation of $\lambda$ can merge two min-max segments.
    \item For a large $\lambda$, we have some long segments due to the intrinsic structure of signal. We need a large increase of $\lambda$ to merge two min-max segments.
\end{itemize}

In consequence, the small values of $\lambda$ (named \emph{denoising regime}) vanish the extremums introduced by noise in averaging over a fine cutting, and the large values of $\lambda$ (named \emph{destructing regime}) regroup the intrinsic extremums of signal. The above analysis shows a specific structure of $g(\lambda)$:  $g(\lambda)$ decreases quickly in denoising regime, but slowly in destruction regime. In the transitory region between two regimes, the noises are nearly all removed, but the shape of original signal is preserved. 
To sum up, a good $\lambda$ may correspond to the discontinuity of the \emph{tendency} (or gradient) of $g(\lambda)$'s decreasing.

\begin{figure}[!htb]
\centering
\subfloat{
\includegraphics[width=0.7\textwidth]{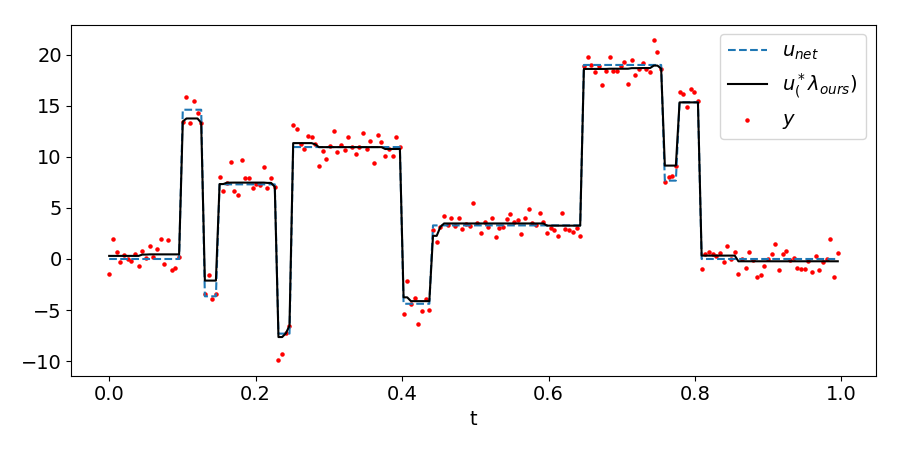}
\label{fig:u_net}
}
\\
\subfloat{
\includegraphics[width=0.7\textwidth]{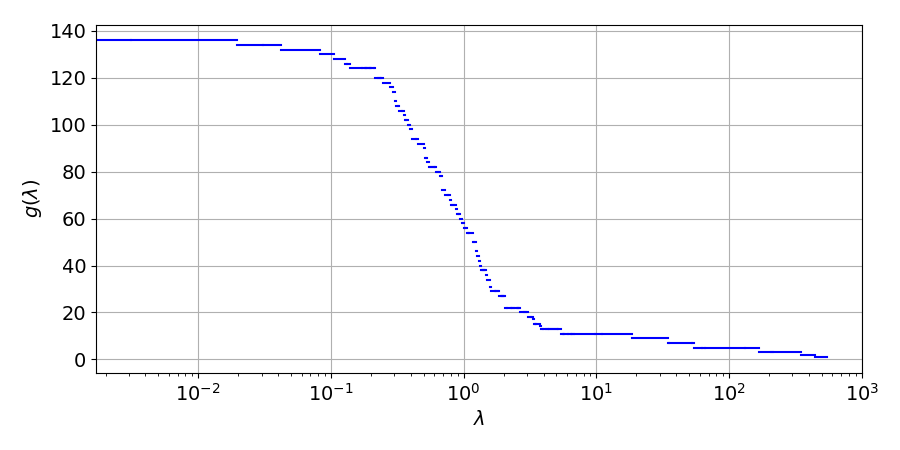}
\label{fig:g_l}
}
\\
\subfloat{
\includegraphics[width=0.7\textwidth]{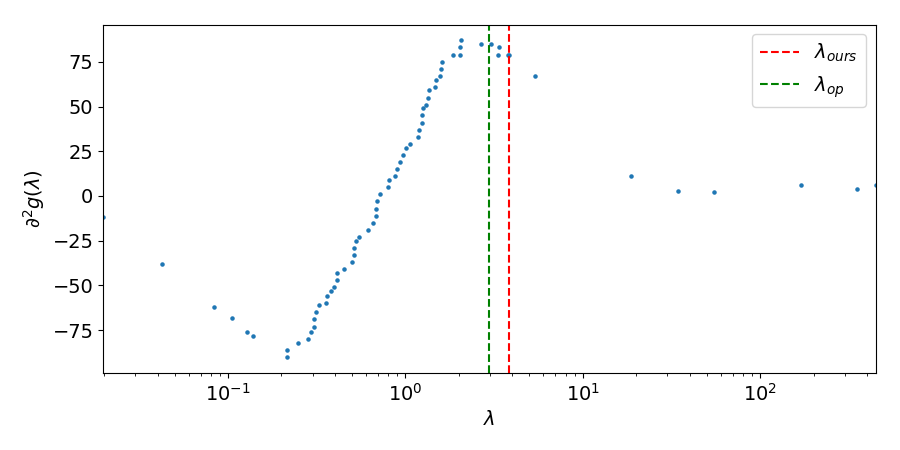}
\label{fig:g_2_l}
}
\\
\caption{Simulation: how to choose a good $\lambda$ from the extremums number ($g(\lambda)$). Top: Simulated block signal $y = u_{net} + \epsilon$ with $\epsilon \sim \mathcal{N}(0, 1)$, SNR = $16.91\mathrm{dB}$, and the restoration with our proposition $u^*(\lambda_{ours})$ is shown in black;  Middle: the extremums number ($g(\lambda)$) in function of $\lambda$; Bottom: $\partial^2 g(\lambda)$  with $\log_{10}(q)=1$ for every $\lambda \in \Lambda^g$. Color: green ($\lambda_{op} = \arg\min \mathcal{R}(\lambda)$), red (our approach).}
\label{fig:simulation}
\end{figure}

We present a simulation of block signal \cite{DonohoDavidL1995AtUS} illustrated in Figure \ref{fig:u_net}: $y$ is a piece-wise constant signal (noted $u_{net}$) adding a Gaussian noise $\epsilon \sim \mathcal{N}(0, 1)$. $g(\lambda)$ of this signal is shown in Figure \ref{fig:g_l}. The shape of $g(\lambda)$ corresponds well to our analysis: 
\begin{itemize}
    \item Denoising regime: for $\lambda< 2$, $g(\lambda)$ decreases quickly since some small min-max segments are merged.
    \item Destructing regime: for $\lambda >7$, $g(\lambda)$ decreases slowly. For $\lambda >500$, the total variation regularisation becomes too important, and $g(\lambda)$ decreases to 1: $u^*(\lambda)$ has only one segment with $u^*(\lambda) = \text{mean}(y)$. 
\end{itemize}

Our intuition is the following: a good $\lambda$ must lie on the last part of transitory region between desnoising regime and the destructing regime, and the transitory region corresponds to the tendency discontinuity of $g(\lambda)$ in function of $\log(\lambda)$ behind a shape decreasing.

Since $g(\lambda)$ is monotonically decreasing and piece-wise constant, the estimation of the tendency of $g(\lambda)$ in function of $\log(\lambda)$ is not obvious. We use the following approximations: for $\lambda\in \Lambda^g$, we introduce the numerical differential operators:
\begin{equation}
    \partial g(\lambda)^+ = g(q \lambda) - g(\lambda)
    \label{equ:right}
\end{equation}
\begin{equation}
    \partial g(\lambda)^- = g(\lambda) - g(\lambda/q)
    \label{equ:left}
\end{equation}
with $q>1$. $\partial g(\lambda)^-$ and $\partial g(\lambda)^+$ are respectively the approximation of the left and right derivative of $g(\lambda)$ in function of $\log(\lambda)$. 

The approximation of second derivative of $g(\lambda)$ is given by: 
\begin{equation}
    \partial^2 g(\lambda) = \partial g(\lambda)^+ - \partial g(\lambda)^-
    \label{equ:d2}
\end{equation}
$\partial^2 g(\lambda)$ with $\log_{10}(q) = 1$ for every $\lambda \in \Lambda^g$ is shown in Figure \ref{fig:g_2_l}.

We propose the following method to estimate a good $\lambda$:
\begin{itemize}
    \item Get $\partial^2 g(\lambda)$ by (\ref{equ:d2}) for $\forall \lambda \in \Lambda^g$, and the transitory region is given by: 
    \begin{equation}
        \lambda_{trans} = \arg \max \partial^2 g(\lambda)
    \end{equation}
    
    \item Adjustment: the transitory region has a similar value of $\partial^2 g(\lambda)$, and $\partial^2 g(\lambda)$ decreases sharply in destruction regime. A good estimation of $\lambda$ can be given by the last $\lambda \in \Lambda^g$ of transitory region. One proposition is to  find out the first sharp decreasing of $\partial^2 g(\lambda)$ for all $\lambda \in \{\lambda \in \Lambda^g\} \cap \{\lambda \geq \lambda_{trans}\}$. We propose to estimate the choice of $\lambda$ by:
        \begin{equation}
            \lambda_{ours} =  \arg \min_{\lambda \in \{\lambda \in \Lambda^g\} \cap \{\lambda \geq \lambda_{trans}\}} \partial^4 g(\lambda)
            \label{equ:generic}
        \end{equation}
         with:
        \begin{equation}
            \partial^4 g(\lambda^g_i) = \partial^2 g(\lambda^g_{i+2}) - 2\partial^2 g(\lambda^g_{i+1}) + \partial^2 g(\lambda^g_{i})
        \end{equation}

\end{itemize}
 The value of $q$ needs to be large in order to avoid the local variation of the gradient of $g(\lambda)$. Typically, $0.5\leq \log_{10}(q) \leq 1$ can well approximate the tendency of $g(\lambda)$. Besides, $q$ can be chosen automatically: we propose $q$ as the length of a long step of $g(\lambda)$ (in logarithm), and our proposition is $q= \max(\log_{10}(\lambda^g_{i+1}/\lambda^g_{i}))$ without the two first steps of $g(\lambda)$.
 
 The complexity for estimating $\lambda_{ours}$ from $g(\lambda)$ is in $O(n)$.

\subsection{Online implementation of DP-TV}
Our objective is to restore a huge amount of noised signals in a real time context. It asks for a weak temporal and spatial complexity for estimating the restoration $u^*(\lambda_{ours})$. In this section, based on the local modification introduced by the new sample, we will propose an online implementation of Algorithm \ref{algo:Lambda_2}.

We note $(\Lambda^{\circ,n}, \Delta g^{\circ,n})$, a set of two vectors of size $n-1$, the solutions of Algorithm \ref{algo:Lambda_2} with $n$ samples. With the new sample $(y_{n+1}, t_{n+1})$, the new results $(\Lambda^{\circ,n+1}, \Delta g^{\circ,n+1})$ based on $n+1$ samples can be obtained by updating $(\Lambda^{\circ,n}, \Delta g^{\circ,n})$. We will only talk about the online estimation of $\Lambda^{\circ, n+1}$ from $\Lambda^{\circ, n}$ in detail. In the following, we note an application of Algorithm \ref{algo:Lambda_2} to a given sequence of signal $(\{y\}, \{\tau\})$ as $\Lambda^\circ = \text{DP-TV}(\{y\}, \{\tau\})$.

After choosing a value of $\lambda$, noted $\hat\lambda$, we can get the restoration $u^*(\hat \lambda)$ for the first $n$ points. Following Theorem \ref{thm:diff_bloque}, the introduction of the new sample changes only the non-isolated sequence of $u^*(\hat \lambda)$. Besides, Proposition \ref{pro:online} implies that $\lambda^{\circ,n}_i = \lambda^{\circ,n+1}_i$ if $\lambda^{\circ,n}_i \leq \hat \lambda$ for the isolated sequence, and the merge of the new restored signal $\hat u (\hat \lambda)$'s segments (i.e $\{\lambda^{\circ,n+1}_i>\hat \lambda\}$) depends on the value of $y_{n+1}$.

\begin{figure}[!htb]
\centering
\begin{tikzpicture}
\centering
\draw[thick,->] (0,0) -- (0.4,1);
\draw[thick,->] (0,0) -- (0.4,-1);
\node at (-0.8,0){$\Lambda^{\circ,n+1}$ };
\node at (3.6,1){$\{\lambda^{\circ,n+1}_i>\hat \lambda\}$: need to update (part 1)};
\node [align=left] at (1.6,-1.2){$\{\lambda^{\circ,n+1}_i\leq\hat \lambda\}$:\\get $u^*(\hat\lambda)$};
\draw[thick,->] (2.9,-1.2) -- (3.3,-0.2);
\draw[thick,->] (2.9,-1.2) -- (3.3,-2.2);
\node [align=left] at (5.4,-0.2){Non-isolated sequence:\\need to update (part 2)};
\node [align=left] at (5.1,-2.2){Isolated sequence:\\unchanged (part 3)};

\end{tikzpicture}
\caption{Illustration of online implementation: the changed and unchanged part of $\Lambda^{\circ,n+1}$ with the introduction of the new sample.}
\label{fig:online_rep}
\end{figure}
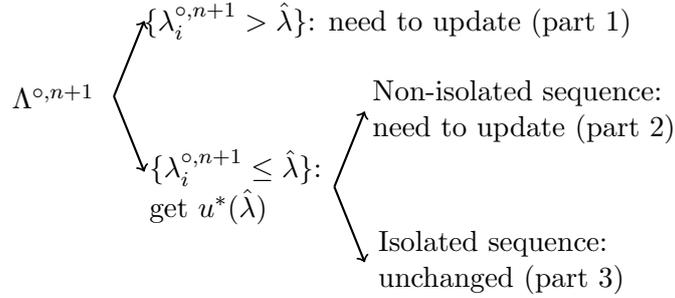

$\hat\lambda$ is indeed a cutting point of $\Lambda^{\circ,n+1}$: $\{\lambda^{\circ,n+1} \leq \hat \lambda\}$ and  $\{\lambda^{\circ,n+1} > \hat \lambda\}$ will be treated separately. We note $m$ and $j^*$ respectively the first point and the first segment of the non-isolated sequence of $u^*(\hat\lambda)$ following Definition \ref{def:non-iso}. The results based on the first $n$ points can be cut into three parts, illustrated in Figure \ref{fig:online_rep}, and we will detail the update procedure for each part.

We treat at first the unchanged part of $\Lambda^{\circ,n+1}$ (part 3 of Figure \ref{fig:online_rep}): all $\lambda^{\circ,n+1}_i \leq\hat{ \lambda}$ remain the same for $i<m$. Formally, we have the following equation:
\begin{equation}
     \lambda^{\circ,n+1}_{\{\lambda^{\circ,n+1}_{1, \cdots, m-1}  \leq \hat{\lambda}  \}} = \lambda^{\circ,n}_{\{\lambda^{\circ,n}_{1, \cdots, m-1}  \leq \hat{\lambda}   \}} 
\end{equation}

For the non-isolated sequence (part 2 of Figure \ref{fig:online_rep}), $\Lambda^a = \lambda^{\circ,n+1}_{\{\lambda^{\circ,n+1}_{m, \cdots, n}  \leq \hat{\lambda} \}}$ can be estimated by $\text{DP-TV}(y^+_{\{m,\cdots,n+1\}},\tau^+_{\{m,\cdots,n+1\}} )$ where:
\begin{itemize}
    \item $y^+_{\{m,\cdots,n+1\}} = \{v^*_{j^*}(\hat\lambda) - \frac{\hat\lambda + \epsilon_\lambda}{2\text{sign}(v_{j^*}-v_{j^*-1})}, y_{\{m,\cdots,n+1\}}\}$
    \item$\tau^+_{\{m,\cdots,n+1\}} = \{1, \tau_{\{m,\cdots,n+1\}}\}$
\end{itemize}
with $\epsilon_\lambda >0$.

The non-isolated and isolated sequences can be assembled into $\Lambda^{temp} = \{\lambda^{temp}_1, \cdots, \lambda^{temp}_{n}\}$ with:
\begin{equation}
\lambda^{temp}_i=\begin{cases}
\lambda^{\circ, n}_i, &  i<m.\\
\lambda^a_{i-m+2} , & i\geq m.\\
\end{cases}
\label{equ:Lambda_n+1_temp}
\end{equation}

It remains the update of $\{\lambda^{\circ,n+1}_i >\hat \lambda\} = \{\lambda^{temp}_i >\hat \lambda\}$ (part 1 of Figure \ref{fig:online_rep}). Let $b =  \{\lambda^{\circ,n+1} > \hat{\lambda}\}$ containing indeed all the junctions of $\hat{u}(\hat\lambda)$'s segments, we can get $\Lambda^b=\lambda^{\circ,n+1}_b = \text{DP-TV}(\hat{v}(\hat{\lambda}), \hat{\mathcal{T}}(\hat{\lambda}))$. 

Finally, $\Lambda^{\circ, n+1}$ can be assembled in the following way:
\begin{equation}
\lambda^{\circ, n+1}_i=\begin{cases}
\lambda^{temp}_i, & \text{if } \lambda^{temp}_i\leq \hat{\lambda}. \\
\lambda^b_{p(i)} & \text{if } \lambda^{temp}_i >\hat{\lambda}.
\end{cases}
\label{equ:Lambda_n+1}
\end{equation}
with $p: \mathbb{Z} \to \mathbb{Z}$ giving the index of $i$ in the vector $b$. 

The other solution vector $\Delta g^{\circ,n+1}$ can be estimated from $\Delta g^{\circ,n}$ in the same way as $\Lambda^{\circ,n+1}$ (from $\Lambda^{\circ,n}$). 

The online version of DP-TV is summed up in Algorithm \ref{algo:online}: DP-TV is applied two times on a small part of the data, and the results are merged following  (\ref{equ:Lambda_n+1}). Concerning the complexity, only a small part of $\Lambda ^{\circ, n+1}$ needs to be recalculated: we obtain $u^*(\hat\lambda)$ in $O(n)$, $\Lambda^a$ in $O((n-m)\log(n-m))$ and $\Lambda^b$ in $O(l_b\log l_b)$ with $l_b = \text{Card}(\Lambda^b)$. The overall complexity depends on the cutting point $\hat\lambda$, a trade-off between the complexity of $\Lambda_a$ and $\Lambda_b$:
\begin{itemize}
    \item A large value of $\hat\lambda$ makes $u^*$ less fluctuating, and the non-isolated sequence may be long, even as long as $y$, which means $m = 1$. In this case, the computation of $\Lambda_a$ is nearly $O(n\log n)$, the same as the offline implementation.
    \item The reconstruction with small value of $\hat\lambda$ has a little non-isolated sequence. However, all $\lambda >\hat\lambda$ with $\lambda \in \Lambda^{\circ, n}$, and need to be recalculated. In the worst case, all the elements of $\Lambda^{\circ, n}$ need to be recalculated in $O(n\log n)$ 
\end{itemize}

A good choice of $\hat\lambda$ may provide a short non-isolated sequence and a little size of $\Lambda_b$, which means $(n-m)\log (n-m) << n$ and $l_b\log (l_b) << n$. In this case, the complexity of the online implementation is $O(n)$. We propose to use the value estimated by the deterministic approach $\hat\lambda = \lambda_{ours}$ or slightly larger than $\lambda_{ours}$ (i.e $\hat\lambda = 2\lambda_{ours}$). See more details in Section \ref{sub:time}.

For estimating $g(\lambda)$ in $O(n)$, we propose to save $\Delta g(\lambda)$ following the order of $\Lambda$. For the online implementation, we believe that the computation of $\Delta g(\lambda)$ for $n+1$ points and $\Lambda^{n+1}$, the ordered list of $\Lambda^{\circ,n+1}$, can been seen as an insertion of a small ordered list ($\Lambda$ of part 2 in Figure \ref{fig:online_rep}) into a large ordered list ($\Lambda$ of part 3 in Figure \ref{fig:online_rep}). After insertion, we need to simply concatenate the new list with $\Lambda$ of part 1 for the final result of $\Lambda^{n+1}$.

\begin{algorithm}
\caption{Online implementation of DP-TV}
\begin{algorithmic} 
\Require $(y_1, \cdots, y_n), (\tau_1,\cdots, \tau_n), \ (y_{n+1}, \tau_{n+1})$
\Require $\Lambda^{\circ,n}, \Delta g^{\circ,n}, \hat{\lambda}$
\State Find non-isolated sequence $(m,\cdots,n)$ of $u^*(\hat\lambda)$
\State ($\Lambda^a, v^a, \Delta g^a) = \text{DP-TV}(y^+_{\{m,\cdots,n+1\}}, \tau^+_{\{m,\cdots,n+1\}})$
\State $\Lambda^{\circ,n+1}=(\lambda^{\circ,n+1}_1, \cdots, \lambda^{\circ,n+1}_{n+1}) = \{\Lambda^{\circ,n}_{\{1,\cdots,m-1\}}, \Lambda^a_{\{2,\cdots\}}\}$
\State  $\Delta g^{\circ,n+1}= \{\Delta g^{\circ,n}_{\{1,\cdots,m-1\}}, \Delta g^a_{\{2,\cdots\}}\}$
\State $b =  \{\lambda^{\circ,n+1} > \hat{\lambda} \}$
\State $(\lambda^{\circ,n+1}_b,\Delta g^{\circ,n+1}_b)  = \text{DP-TV}(\hat{u}(\hat{\lambda}), \hat{\mathcal{T}}(\hat{\lambda}))$
\Return $\Lambda^{\circ,n+1}$ and $\Delta g^{\circ,n+1}$ 

\end{algorithmic}
\label{algo:online}
\end{algorithm}

\section{Applications}
\label{sec:res}
We have proposed an automatic TV-restoration method for the real time context with limited computation resource. In this section, we will evaluate the restoration performance and the execution time with some simulated signal, and show an application with some real data collected from a Saint-Gobain's plant.

\subsection{Restoration performance evaluation}
In the section, we evaluate the restoration with $\lambda_{ours}$ proposed by our method (\ref{equ:generic}) and compare with different existent methods.
\subsubsection{Metrics and criteria}
 We use the mean square error $\mathcal{R}(\lambda)$ (\ref{equ:mse}) between the restored signal and the original signal to evaluate a candidate of $\lambda$. The optimal value $\lambda_{op}$ is given by (\ref{equ:l_op}).
For comparing the performance between two given values $\lambda_1$ and $\lambda_2$, we apply the following criteria:
\begin{equation}
    d(\lambda_1, \lambda_2) = \mathcal{R}(\lambda_1) - \mathcal{R}(\lambda_ 2) 
\end{equation}
$d(\lambda, \lambda_{op})\geq 0, \forall \lambda\in \mathbb{R}^+$, since $\lambda_{op}$ provides the minimum of $\mathcal{R}(\lambda)$. The smaller $d(\lambda, \lambda_{op})$ is, the better the estimation $\lambda$ is.

However, it is not clear how to fix a threshold over $d(\lambda, \lambda_{op})$ between a good and unacceptable estimation of $\lambda$. It depends on the variance of noise and the shape of the original signal. An alternative way is to compare with the estimation with K-fold cross-validation. 
\subsubsection{Cross-validation}
The objective of cross validation is to assess how a model behaves on an independent data set. By selecting the best model, we get an estimation of hyper-parameter. If $d(\lambda, \lambda_{op})$ is near $d(\lambda_{cv}, \lambda_{op})$ with $\lambda_{cv}$ estimated by cross-validation and $\lambda$ estimated by another method, we can say this method has a similar performance as cross-validation.

We show how we use cross-validation (CV) in 1D total variation denoising. The observed signal $\{y_i,t_i\}_{1,\cdots, n}$ is splitted randomly into $k$ folds with (nearly) equal size in each fold, noted $(\mathbf{y}^{i}, \mathbf{t}^{i})$ for the $i^{th}$ fold and $m_{i} = \text{Card}(\mathbf{y}^{i})$. 
A model is estimated without a fold of data. For example, the model without $i^{th}$ fold is noted:
\begin{equation}
    \mathbf{u}^{-i}(\lambda) = \arg\min_u \{F(\mathbf{y}^{-i}, u, \mathbf{t}^{-i}, \lambda)\} 
\end{equation}
where $(\mathbf{y}^{-i}, \mathbf{t}^{-i})$ the observed signal without $i^{th}$ fold.

The prediction $\hat{u}^{-i}(t)$ at time $t$ is given by the linear interpolation of ($\mathbf{u}^{-i}(\lambda), \mathbf{t}^{-i}$). The empirical error of $\mathbf{u}^{-i}(\lambda)$ is estimated over $i^{th}$ fold which does not participate into the construction of the model:
\begin{equation}
    e^{-i}(\lambda) = \frac{1}{m_i}\sum_{(y,t)\in (\mathbf{y}^{i}, \mathbf{t}^{i})} (y-\hat{u}^{-i}(t))^2 
\end{equation}

Finally, we can estimate the hyper-parameter $\lambda$ by:
\begin{equation}
    \lambda_{cv} = \arg\min_{\lambda \in \mathbb{R}^+} \sum_{i=1}^K e^{-i}(\lambda)
    \label{equ:cv_est}
\end{equation}

\begin{theoremEnd}[no proof end]{req} 
A sequence of signal can be seen as a time series. The random split of training and validation sets is not appropriated for time series \cite{arlot2010survey}, since training and validation set are not independent anymore. However, we are interested in the performance of our model in training data, but not in an independent data set, so the random split of data is justified. 
\label{rq:cv}
\end{theoremEnd}
\begin{figure}[!htb]
\centering
\subfloat{
\includegraphics[width=0.48\textwidth]{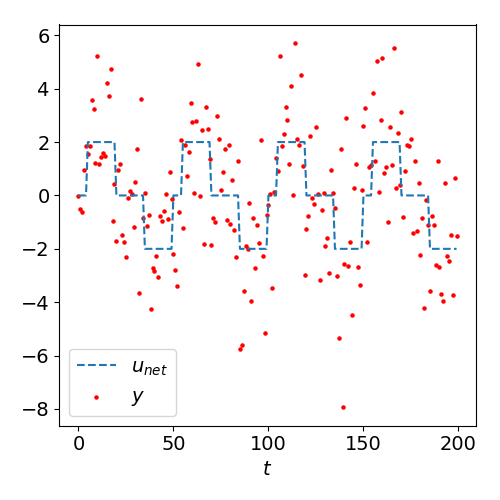}
\label{fig:s_1}
}
\subfloat{
\includegraphics[width=0.48\textwidth]{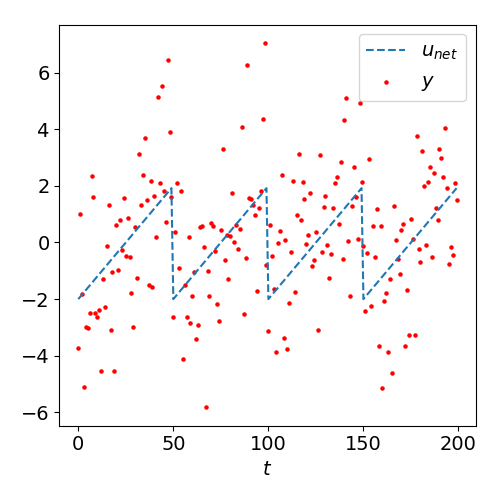}
\label{fig:s_2}
}
\caption{2 types of simulated periodic signal ($u_{net}$). Examples of $y= u_{net} + \epsilon$ with $\epsilon\sim\mathcal{N}(0, 4)$ are shown. Left: Piece-wise constant; Right: Piece-wise linear.}
\label{fig:signal}
\end{figure}

\subsubsection{Results}
We are interested in 2 types of periodic signal ($u_{net}$) shown in Figure \ref{fig:signal}: piece-wise constant and piece-wise linear. Gaussian noise $\epsilon\sim\mathcal{N}(0, \sigma^2)$ is added to $u_{net}$. Examples of simulated noised signals are also shown in Figure \ref{fig:signal}. Now, we compare our proposition, $\lambda = \lambda_{ours}$, estimated by (\ref{equ:generic}) using $\log_{10} q \in \{0.5,1\}$ and $q$ automatic, with the proposition by 10-fold cross-validation (\ref{equ:cv_est}). For each type of signal and each $\sigma \in \{0.5, 1, 1.5, 2, 2.5, 3\}$, 500 simulations are done. 
\begin{figure}[!htb]
\centering
\subfloat[Piece-wise constant]{
\includegraphics[width=12cm]{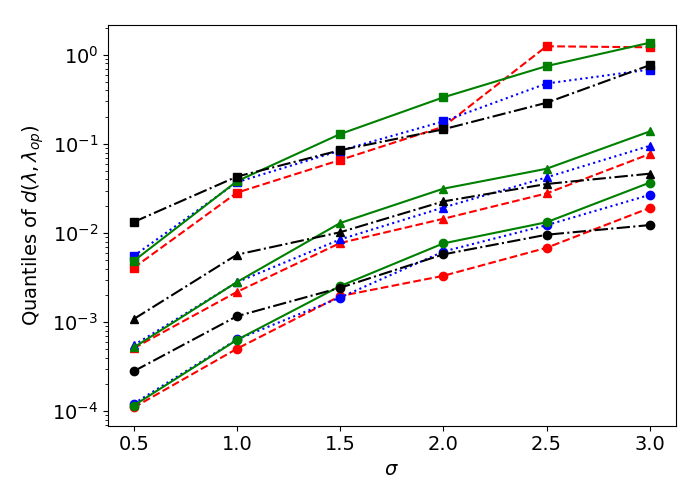}
\label{fig:pwc_1}
}
\quad
\subfloat[Piece-wise linear]{
\includegraphics[width=12cm]{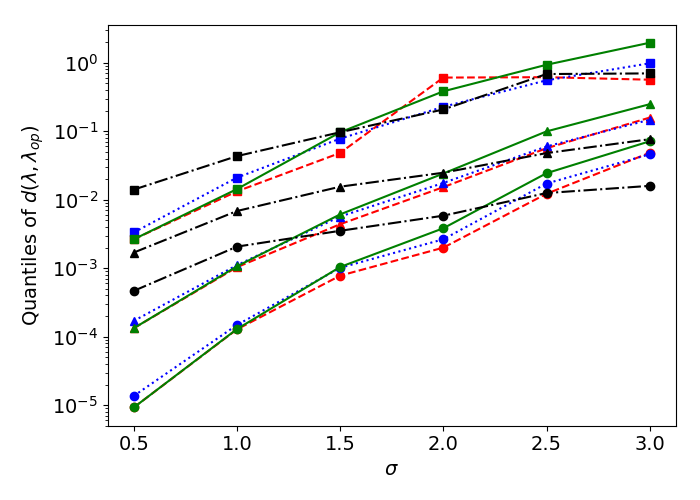}
\label{fig:pwl_1}
}
\\
\caption{$d(\lambda, \lambda_{op})$'s statistical characteristics of 500 simulations for 2 types of periodic signal with different Gaussian noise variance. We have tested our approach with $\log_{10}(q) = 0.5, 1$ and automatic choice of $q$. Solid line (green): automatic choice of $q$;
 Dots line (blue): $\log_{10}(q) = 0.5$; Dashed line (red): $\log_{10}(q) = 1$; Dash-dot line (black): 10-fold cross-validation. Circle marker: $25\%$ quantile; Triangle marker: median; Square marker: $95\%$ quantile.}

\label{fig:box_simu}
\end{figure}

The statistical characteristics of $d(\lambda_{ours}, \lambda_{op})$ with different parameters $q$ and $d(\lambda_{cv}, \lambda_{op})$ are shown in Figure \ref{fig:box_simu}. For low noise variance, all those methods provide a reconstruction $u^*(\lambda)$ nearly identical to $u^*(\lambda_{op})$. With the growing noise variance, the estimation of $\lambda$ is more and more difficult, and the performance of all methods decreases. For high noise variance (i.e. $\sigma = 3$), those methods still provide a similar restoration as $u ^*(\lambda_{op})$ for $50\%$ cases with $d(\lambda, \lambda_{op})<0.1$.

Our methods with different parameters have a similar restoration performance as cross-validation for both types of periodic signal.

\subsubsection{Comparison with existing methods}

We compared our approach with some existing methods: Stein unbiased risk estimation (SURE) \cite{stein1981estimation} and Adaptive universal threshold (AUT) \cite{sardy2016threshold}. Both methods are based on the known noise variance ($\sigma^2$). SURE assesses a criteria for several candidates $\lambda$, while AUT calculates $u^*(\lambda)$ for only two nice choices of $\lambda$. By following, we will compare those methods with $\sigma$ known or unknown. In the case of unknown, the noise variance is estimated by \cite{donoho1994ideal}. 

\begin{figure}[!htb]
\centering
\subfloat{
\includegraphics[width=\textwidth]{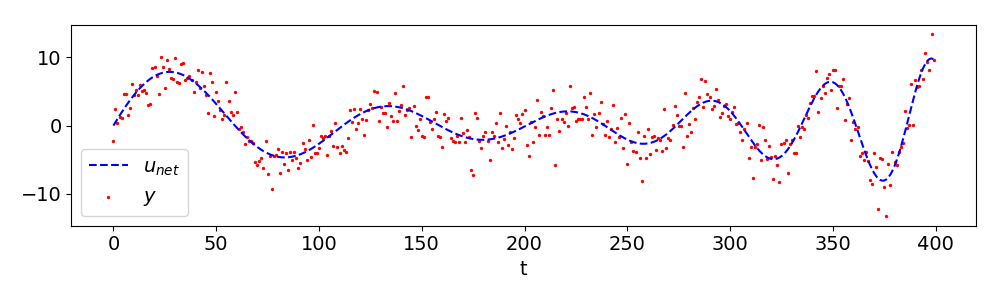}
}
\caption{Example of non-periodic simulated signal. $u_{net}$: original signal; $y$: noised signal. $y = u_{net}+\epsilon$ with $\epsilon\sim \mathcal{N}(0,4).$}
\label{fig:np_signal}
\end{figure}

At first, we applied those methods to restore a non-periodic signal, shown in Figure \ref{fig:np_signal}, with different types of noise. The average value of $\mathcal{R}(\lambda)$ of different methods over 500 simulations are summarized in Table \ref{tab:soa}. The results show that our approaches with $q\in \{0.5, 0.75, 1\}$ and $q$ automatic provide a similar performance as SURE for different types of noise, and out-perform AUT in the strong noise cases.

The second signal we tested is block signal (c.f Figure \ref{fig:u_net}) with different sampling points (frequency). The average values of $\mathcal{R}(\lambda)$ with $n=199$, 499 and 999 are shown in Table \ref{tab:soa_block}. The performance of all the methods improves as the increasing of the sampling points. AUT and SURE have a smaller error than our approaches, but the difference between ours and the other approaches remains around the order of magnitude of $0.01$. Indeed, the performance of our approaches based on the extremums number depends on the sampling frequency: average over a big number of points allows a better restoration of the intrinsic extremums of the original signal.

Concerning the complexity for estimating the choice of $\lambda$ and the corresponding solution $u^*(\lambda)$, in applying a solution algorithm in $O(n\log n)$ (e.g. ours), SURE can estimate a choice of $\lambda$ in $O(n\log n + Nn)$ with $N$ the number of candidates and $O(Nn)$ the complexity for estimating $u^*(\lambda)$ from $\Lambda^\circ$ for all candidates, while AUT and our method are in $O(n\log n)$. For the online implementation, the complexity of SURE is in $O(n + Nn)$, while AUT and our approach are in $O(n)$. In the case of huge number of candidates, SURE is slower than AUT and our method.

We can not compare our approach with the heuristic method proposed in \cite{pollak2005nonlinear} due to the lack of implementation details. This method tracks the variation of $|u^*(\lambda_{l-1}) - u^*(\lambda_{l})|^2_2$ for every elements of $\Lambda$. However, the estimation of all $u^*(\lambda)$ for every $\lambda \in \Lambda$ is in $O(n^2)$, and this method may not be appropriated for the real time context.

In summary, our approach works well for the signal restoration under different types of noise with high sampling frequency.
\begin{table}[!htb]
    \centering
    \begin{tabular}{l|ccccc}


    \hline
 &  $\mathcal{N}(0,1)$  & $\mathcal{N}(0,2^2)$& $\mathcal{N}(0,3^2)$ &  $\mathcal{N}(0,2^2)$\\
 &&&&$+ \mathcal{U}[-2,2]$ \\
\hline
$\min \mathcal{R}(\lambda)$ & 16.16 &41.87 & 73.04 & 51.59\\
\hline\hline
\multicolumn{5}{c}{$\sigma$ known}\\
         \hline
AUT & 16.43 &58.87 & 160.28& 88.31\\
         \hline
  SURE     & 17.07& 45.31& 81.67& 56.39\\
  \hline\hline
\multicolumn{5}{c}{$\sigma$ unknown, $\hat \sigma$ estimated by \cite{donoho1994ideal}}\\
          \hline
AUT & 16.52& 60.17&160.11&90.51\\
         \hline
  SURE     & 17.22& 46.44& 85.02& 57.82\\

  \hline\hline
    \multicolumn{5}{c}{Our approach}\\
  \hline
 $\log_{10}(q) = 0.5$ & 16.64& 44.60& 82.44&55.38\\
  \hline
 $\log_{10}(q) = 0.75$ & 16.58& 44.04& 80.26&54.41\\
  \hline
 $\log_{10}(q) = 1$ & 16.58& 44.30& 83.95&55.67\\

  \hline
  $q$ automatic & 16.58&45.02&87.64&56.77\\
  \hline
    \end{tabular}
    \caption{The average value of $\mathcal{R}(\lambda)*100$ of 500 simulations with a non-periodic signal shown in Figure \ref{fig:np_signal} with different types of noise (Gaussian and uniform) shown in the first line of table.}
    \label{tab:soa}
\end{table}
\begin{table}[!htb]
    \centering
    \begin{tabular}{l|cccc}


    \hline
 & 199  & 499& 999\\
\hline
$\min \mathcal{R}(\lambda)$ & 23.30 &11.49& 6.42\\
\hline\hline
\multicolumn{4}{c}{$\sigma$ known}\\
         \hline
AUT & 25.01 &11.92& 6.56\\
         \hline
  SURE     & 24.97&12.24&6.83 \\
   \hline\hline
\multicolumn{4}{c}{$\sigma$ unknown, $\hat \sigma$ estimated by \cite{donoho1994ideal}}\\
          \hline
      AUT & 26.34&12.08&6.59\\
         \hline
  SURE     & 25.26&12.42&6.84\\

  \hline\hline
      \multicolumn{4}{c}{Our approach}\\

  \hline
 $\log_{10}(q) = 0.5$ & 27.88& 13.36& 7.75\\
  \hline
 $\log_{10}(q) = 0.75$ &28.55& 13.57& 7.39\\
  \hline
 $\log_{10}(q) = 1$  & 30.17& 14.63&7.72\\
 \hline 
  $q$ automatic & 30.54&13.52&7.51\\
  \hline
    \end{tabular}
    \caption{The average value of $\mathcal{R}(\lambda)*100$ of 500 simulations of block signal (SNR=$16.91\mathrm{dB}$) with different sampling points. Examples of original and noised signal are shown in Figure \ref{fig:u_net}.}
    \label{tab:soa_block}
\end{table}

\subsection{Execution time evaluation}
\label{sub:time}
In this section, we compare the execution time of offline and online implementations. We measure the execution time for estimating $\Lambda^\circ$ with DP-TV (offline, $O(n\log n)$) and its online implementation. For the online implementation, the estimation with $n+1$ points is based on the result with $n$ points, while the offline implementations re-estimate from scratch the results. All the results shown are the average over 20 simulations: from $50$ to $500$ sample points of a periodic piece-wise linear whose one period is the same as Figure \ref{fig:s_2}. 

\begin{figure}[!htb]
\centering
\subfloat{
\includegraphics[width=0.6\textwidth]{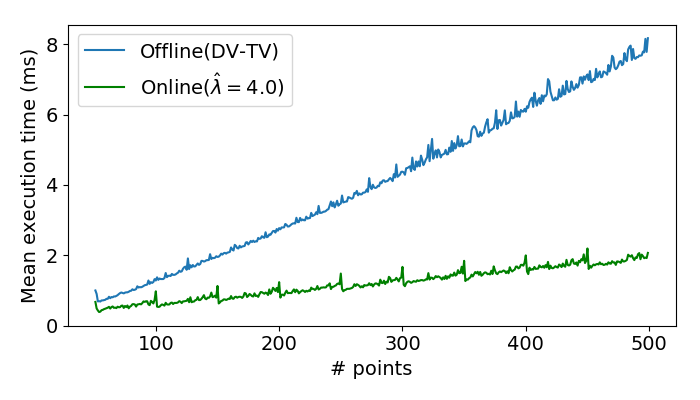}
\label{fig:on_off}
}
\\
\subfloat{
\includegraphics[width=0.6\textwidth]{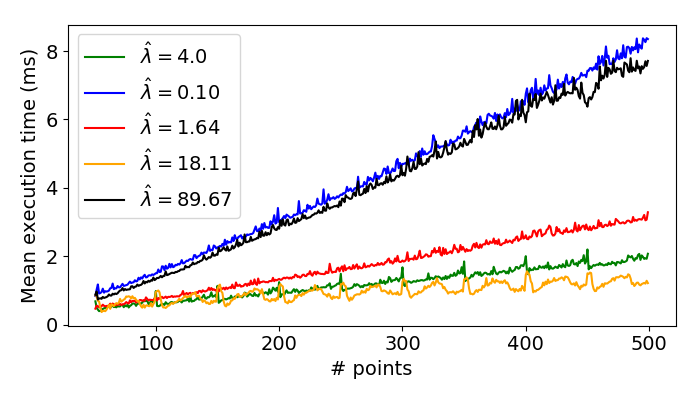}
\label{fig:on}
}
\\
\subfloat{
\includegraphics[width=0.6\textwidth]{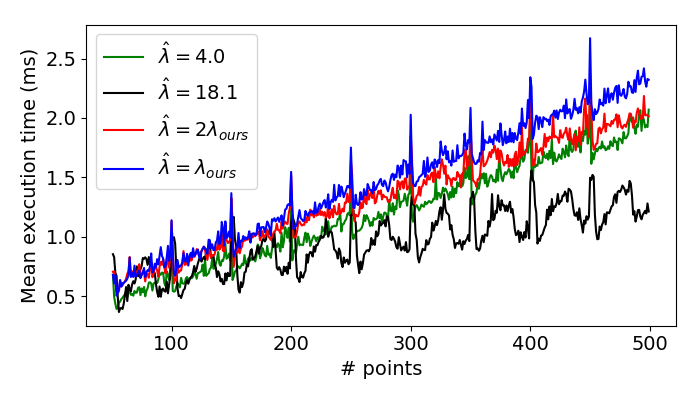}
\label{fig:on_op}
}
\\
\caption{Mean execution time between offline and online implementation with different cutting points $\hat\lambda$. Execution time is averaged over 20 simulations. Top: Online vs offline (DP-TV in $O(n\log n)$); Median: Online implementation with different cutting points $\hat\lambda$; Bottom: A proposition of the choice of cutting point: for the $i^{th}$ point, $\hat\lambda_i = \lambda_{ours}^{i-1}$ with $\lambda^{i-1}_{ours}$ our hyper-parameter estimation based on the first $i-1$ points with the automatic choice of $q$.}

\label{fig:on_off_all}
\end{figure}

The performance of the online implementation depends on the choice of the cutting point $\hat \lambda$. We will, at first, fix $\hat\lambda = 4$, and then discuss about the choice of $\hat \lambda$. The average execution time is shown in Figure \ref{fig:on_off}. Online implementation has a much smaller execution time than offline implementation: online implementation needs less than 2ms for the $500^{th}$ sampling point, while DP-TV needs more than 5ms for a sequence of 500 points.

We compare some choices of the cutting point $\hat\lambda$, and the results are shown in Figure \ref{fig:on}. For too small and too large value of $\hat\lambda$ (i.e. $\hat\lambda=0.1$ and $89.67$), the execution time is nearly the same as the offline method since almost all the elements are included in $\Lambda^b$ for the first case and in $\Lambda^a$ for the second case, which need to be recalculated.
The parameter $\hat\lambda \in \{1.64, 4, 18.44$\}, providing a small execution time, seems to us a good choice of the cutting point: all needs less than 3ms for the $500^{th}$ point.

However, the choice of $\hat\lambda$ is not obvious: it depends on the shape of signal and the noise added. We propose to choose $\hat{\lambda} = \lambda_{ours}$, the value estimated by our algorithm or slightly larger: for the new $i^{th}$ point, we take $\hat{\lambda}_i = \lambda_{ours}^{i-1}$ estimated from $\Lambda^{\circ, i-1}$ (the result of $i-1$ points). It is a natural choice since $\lambda_{ours}^{i-1}$ needs to be calculated for restoring the $i-1$ points signal. The results of this proposition are shown in Figure \ref{fig:on_op}. The results with $\hat\lambda = \lambda_{ours}$ and $2\lambda_{ours}$, containing the time for estimating $\lambda_{ours}$, have a similar execution time as $\hat\lambda = 4$ in our simulation. Our proposition is not optimal, but it still provides a good online performance. We have to remark that the execution time is more important for each $50^{th}$ points since the new period of the original signal begins and the non-isolated sequence is much longer. Once again, the execution time of online implementation depends on the shape of the signal.

\subsection{Real data application}
\begin{figure}[!htb]
\centering
\subfloat{
\includegraphics[width=0.9\textwidth]{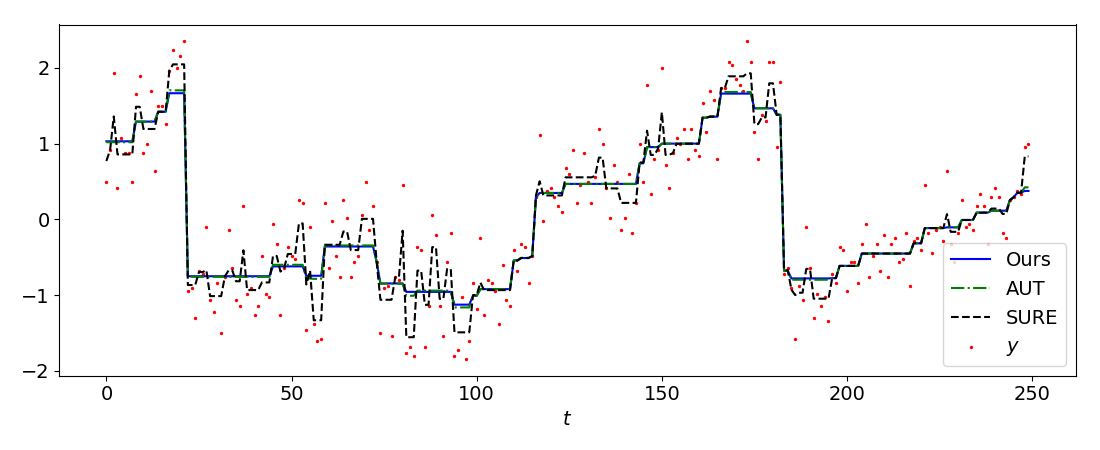}
}
\caption{Real signal example: TV-denoising with our approach (blue), AUT (green) and SURE (black).}
\label{fig:tpc}
\end{figure}

The high performance of our method encourages the application in real data. We show, in Figure \ref{fig:tpc}, an example of signal collected from a plant of Saint-Gobain and the restoration proposed by our approach, AUT and SURE. The original signal is unknown for the real data, so we can not compare qualitatively different methods. An alternative way is to validate the restoration by industrial process engineer. The proposition of SURE, containing some peaks, is irregular, while the solutions proposed by our approach and AUT restore well the variations of measured signal. The process engineer confirms the restoration proposed by our approach is well fitted for the further applications. 

\section{Conclusions and perspectives}
\label{sec:concl}
In this paper, we analysed the behavior of Total Variation restored signal in function of the hyper-parameter $\lambda$ and the introduction of a new sample at the end of the sequence. We propose different algorithms for the real time automatic TV-denoising: 
\begin{itemize}
    \item Based on \cite{Tibshirani} and \cite{pollak2005nonlinear}, we propose some algorithms for estimating efficiently the restored signal $u^*_{TV}$. Our proposition combines the advantages of the existing methods: efficient for hyper-parameter selection as well as online data stream restoration.
    \item We propose a rapid heuristic method for selecting a good choice of $\lambda$ based on the variation of the extremums number of the restored signal in function of $\lambda$. We have compared our methods with some existing methods (cross-validation, AUT and SURE) under Gaussian and/or uniform noise, and the simulations show that our method has a similar performance as cross-validation and SURE which are not compatible with the real time context. The selection of hyper-parameter stays an open question for other noise models.
\end{itemize}

The overall complexity for obtaining a restored signal with an automatic choice of $\lambda$ is in $O(n\log n)$ for the offline implementation and in $O(n)$ for the online implementation in the best case with a space complexity in $O(n)$.


The low time space complexities and the good performance of the choice of $\lambda$ provide a large application field of our methods: for example, monitoring in real time a huge amount of sensors. We believe that there remains some interesting theoretical and practical issues for future research. An important theoretical and practical issue is the adaptation of the presented restoration procedure to low cost embedded devices (typically using 8 bits arithmetic).

\appendix
\section{Proofs}
\label{sec:proofs}
We give, at first, some lemmas used by our proofs. Those lemmas are based on a fixed value of $\lambda$, so $\lambda$ is omitted in most of the notation.
\begin{theoremEnd}[no link to proof]{lem}
	For the $k^{th}$ segment ($\mathcal{N}^*_k$) with $1<k\leq K$, let $\mathcal{T}_{m,k} = \sum_{j=i^{k}_1}^{i^{k}_{m}} \tau_i$ and $\overline{y}_{m,k} = \frac{1}{\mathcal{T}_{m,k}}\sum_{j=i^{k}_1}^{i^{k}_{m}} \tau_j y_j$, we have the following inequalities:
	\begin{itemize}
		\item $\overline{y}_{m,k}\geq v^*_{k}$ for $m = 1,\cdots,n^*_{k} $ if $v^*_{k-1}<v^*_{k}$.
		\item $\overline{y}_{m,k} \leq v^*_{k}$ for $m = 1,\cdots,n^*_K $ if $v^*_{k-1}>v^*_{k}$.
	\end{itemize}
	\label{lem:overline_y}
\end{theoremEnd}
\begin{proofEnd}
	We can easily show that $\sum_{j=i^{k}_1}^{i^{k}_{m}} \tau_j(y_{j}- \overline{y}_{m,k})^2 \leq  \sum_{j=i^{k}_1}^{i^{k}_{m}} \tau_j(y_{j}- v^*_k)^2$.

	For the cases in which $\overline{y}_{m,k}\in [min(v^*_{k-1}, v^*_{k}), max(v^*_{k-1}, v^*_{k})]$, $\sum_{j=i^{k}_1}^{i^{k}_{m}} \tau_j(y_{j}- \overline{y}_{m,k})^2 + \sum_{j=i^{k}_{m+1}}^{i^{k}_{n_{k}}} \tau_j(y_{j}- 
	v^*_{k})^2 + \lambda (v^*_{k} - v^*_{k-1}) < \sum_{j\in \mathcal{N}_{k}} \tau_j(y_{j}-  
	v^*_{k})^2 + \lambda (v^*_{k} - v^*_{k-1})$.

	Since $v^*_{k}$ is given by the minimisation of $F_n$, $\sum_{j=i^{k}_1}^{i^{k}_{m}} \tau_j(y_{j}- \overline{y}_{m,k})^2 + \sum_{j=i^{k}_{m+1}}^{i^{k}_{n_{k}}} \tau_j (y_{j}- 
	v^*_{k})^2 + \lambda (v^*_{k} - v^*_{k-1}) \geq \sum_{j\in \mathcal{N}_{k}} \tau_j(y_{j}-  
	v^*_{k})^2 + \lambda (v^*_{k} - v^*_{k-1})$.
	
	So $\overline{y}_{m,k} \not \in [min(v^*_{k-1}, v^*_{k}), max(v^*_{k-1}, v^*_{k})]$. 
\end{proofEnd}
\begin{theoremEnd}[no link to proof]{lem}
	 For the $k^{th}$ segment ($\mathcal{N}^*_k$) with $1\leq k<K$: let $\mathcal{T}_{-m,k} = \sum_{j=i^{k}_{n_k^*-m}}^{i^{k}_{n_k}} \tau_i$ and $\overline{y}_{-m,k} = \frac{1}{\mathcal{T}_{-m,k}}\sum_{j=i^{k}_{n_k^*-m}}^{i^{k}_{n_k}} \tau_j y_j$, we get:
	\begin{itemize}
		\item $\overline{y}_{-m,k}\geq v^*_{k}$ for $m = 1,\cdots,n_{k} $ if $v^*_{k+1}<v^*_{k}$.
		\item $\overline{y}_{-m,k} \leq v^*_{k}$ for $m = 1,\cdots,n_K $ if $v^*_{k+1}>v^*_{k}$.
	\end{itemize}
	\label{lem:overline_y2}
\end{theoremEnd}
\begin{proofEnd}
Same as Lemma \ref{lem:overline_y}.
\end{proofEnd}
\begin{theoremEnd}[no link to proof]{lem}
\begin{itemize}
    \item If $u^*_m = \hat{u}_m,\ \exists m \leq n$, then $u^*_i = \hat{u}_i$, $\forall i\leq m$.
    \item If $\text{sign}(v^*_{j-1}-v^*_{j}) = \text{sign}(v^*_{j}-\hat{v}_{j})$, $\hat{v}_i = v^*_i$, $\forall i \leq j-1$.
\end{itemize}
\label{lem:v_diff}
\end{theoremEnd}
\begin{proofEnd}
The uniqueness of the solution of (\ref{equ:loss_lions}) is shown in \cite{chambolle1997image}. 

$(u_1,\cdots,u_{m}) = \arg\min F_{m}$ for $m<n$ in condition of $u_m = u^*_m$. When $u^*_m = \hat{u}_m$, two solutions $(u^*_1,\cdots,u^*_{m})$ and $(\hat{u}_1,\cdots, \hat{u}_{m})$ minimize the same functional $F_{m}$ with the same condition $u_m = u^*_m$. So $(u^*_1,\cdots,u^*_{m}) = (\hat{u}_1,\cdots, \hat{u}_{m}) = (u_1,\cdots,u_{m})$.

Assuming that the first $j-1$ segments contains $m$ points, $(v^*_{1},\cdots, v^*_{j-1}) = \arg\min F_m$ under the boundary condition of $u^*_{m} < u^*_{m+1}$. If $\text{sign}(v^*_{j-1}-v^*_{j}) = \text{sign}(v^*_{j}-\hat{v}_{j})$, the boundary condition is unchanged. So $\hat{v}_i = v^*_i$, $\forall i \leq j-1$.
\end{proofEnd}
\begin{theoremEnd}[no link to proof]{lem}
In the case where $v^*_{K^*-1}<v^*_{K^*}$:
\begin{itemize}
    \item 	For $y_{n+1}\geq v^*_{K^*}$, we have $v^*_j = \hat{v}_j$ and  $\mathcal{N}_j^* = \hat{\mathcal{N}}_j$,$\forall j < K^*$.
    
    \item  For $y_{n+1} <v^*_{K^*}$, if $\exists l$ such that $v^*_l > v^* _{l+1}$, then $v_j^* = \hat{v}_j$ and $\mathcal{N}_j^* = \hat{\mathcal{N}}_j$, $\forall j \leq l$
\end{itemize}
\label{lem:bloquant}
\end{theoremEnd}
\begin{proofEnd}
The first point is trivial following Lemma \ref{lem:v_diff}.

For the sake of clarity, we note the last segment together with the new observation as $\mathcal{N}_a = \{\mathcal{N}^*_{K^*}, n+1\}$.
For $y_{n+1} < v^*_{K^*}$, we will discuss the following cases:
\begin{itemize}
    \item Influence on $\mathcal{N}^*_{K^*}$: we have $\hat u_i < u ^*_i, \forall i \in \hat{\mathcal{N}_{K^*}}$ following Lemma \ref{lem:overline_y}.
    \item Influence on $\mathcal{N}^*_{K^*-1}$ with $v^*_{K^*-1} < v^*_{K^*}$: if $\hat v_{K^*} > v^*_{K^*}$, then $\hat v_{K^*-1} =v^*_{K^*-1} $ following Lemma \ref{lem:v_diff}. In the other case, $\{\mathcal{N}^*_{K^*-1}, \mathcal{N}_{a}\}$ may either merge into one segment or split into several, and $\hat v_{K^*-1} < v^*_{K^*-1}$ following Lemma \ref{lem:overline_y}. In other words, if we have a sequence $v^*_j< \cdots< v^*_{K^*}$ with $v^*_{i+1} - v^*_{i}$ small enough for all $i\geq j$, then $y_{n+1}$ can change the value of the elements $\mathcal{N}^*_j$.
    \item Influence on $\mathcal{N}^*_{i}$ with $v^*_{i+1}<\cdots<v^*_{K^*-1} < v^*_{K^*}$ and $v^*_{i} > v^*_{i+1}$: from the previous analysis, we have $\hat v_{i+1} \leq v^*_{i+1}$ regardless of the value of $y_{n+1}$ (under the condition $y_{n+1} < v^*_{K^*}$). Following Lemma \ref{lem:v_diff}, $\hat v_{j} = v^*_{j}$, and $\hat{\mathcal{N}}_j =\mathcal{N}^*_j$ for $j\leq i$.
\end{itemize}

\end{proofEnd}
\printProofs

\section*{Acknowledgment}
The authors would like to thank David Brie (CRAN, Université de Lorraine), Paul Narchi, Diane Bienaimé (Saint-Gobain Research Paris) and Christian Fourel (Saint-Gobain Pont-à-Mousson) for their constructive advices.

This work has been supported by the EIPHI Graduate School (ANR-17-EURE-0002).

\bibliographystyle{IEEEtran}
\bibliography{references}

\begin{thebibliography}{10}
\providecommand{\url}[1]{#1}
\csname url@samestyle\endcsname
\providecommand{\newblock}{\relax}
\providecommand{\bibinfo}[2]{#2}
\providecommand{\BIBentrySTDinterwordspacing}{\spaceskip=0pt\relax}
\providecommand{\BIBentryALTinterwordstretchfactor}{4}
\providecommand{\BIBentryALTinterwordspacing}{\spaceskip=\fontdimen2\font plus
\BIBentryALTinterwordstretchfactor\fontdimen3\font minus
  \fontdimen4\font\relax}
\providecommand{\BIBforeignlanguage}[2]{{%
\expandafter\ifx\csname l@#1\endcsname\relax
\typeout{** WARNING: IEEEtran.bst: No hyphenation pattern has been}%
\typeout{** loaded for the language `#1'. Using the pattern for}%
\typeout{** the default language instead.}%
\else
\language=\csname l@#1\endcsname
\fi
#2}}
\providecommand{\BIBdecl}{\relax}
\BIBdecl

\bibitem{savitzky1964smoothing}
A.~Savitzky and M.~J. Golay, ``Smoothing and differentiation of data by
  simplified least squares procedures.'' \emph{Analytical chemistry}, vol.~36,
  no.~8, pp. 1627--1639, 1964.

\bibitem{Winkler}
G.~Winkler, \emph{Image Analysis, Random Fields and Markov Chain Monte Carlo
  Methods: A Mathematical Introduction (Stochastic Modelling and Applied
  Probability)}.\hskip 1em plus 0.5em minus 0.4em\relax Berlin, Heidelberg:
  Springer-Verlag, 2006.

\bibitem{rudin1992nonlinear}
L.~I. Rudin, S.~Osher, and E.~Fatemi, ``Nonlinear total variation based noise
  removal algorithms,'' \emph{Physica D: nonlinear phenomena}, vol.~60, no.
  1-4, pp. 259--268, 1992.

\bibitem{chambolle1997image}
A.~Chambolle and P.-L. Lions, ``Image recovery via total variation minimization
  and related problems,'' \emph{Numerische Mathematik}, vol.~76, no.~2, pp.
  167--188, 1997.

\bibitem{harchaoui2010multiple}
Z.~Harchaoui and C.~L{\'e}vy-Leduc, ``Multiple change-point estimation with a
  total variation penalty,'' \emph{Journal of the American Statistical
  Association}, vol. 105, no. 492, pp. 1480--1493, 2010.

\bibitem{ortelli2018total}
F.~Ortelli, S.~van~de Geer \emph{et~al.}, ``On the total variation regularized
  estimator over a class of tree graphs,'' \emph{Electronic Journal of
  Statistics}, vol.~12, no.~2, pp. 4517--4570, 2018.

\bibitem{kolmogorov2016total}
V.~Kolmogorov, T.~Pock, and M.~Rolinek, ``Total variation on a tree,''
  \emph{SIAM Journal on Imaging Sciences}, vol.~9, no.~2, pp. 605--636, 2016.

\bibitem{Tibshirani}
\BIBentryALTinterwordspacing
R.~J. Tibshirani and J.~Taylor, ``The solution path of the generalized lasso,''
  \emph{The Annals of Statistics}, vol.~39, no.~3, p. 1335–1371, Jun 2011.
  [Online]. Available: \url{http://dx.doi.org/10.1214/11-AOS878}
\BIBentrySTDinterwordspacing

\bibitem{pollak2005nonlinear}
I.~Pollak, A.~S. Willsky, and Y.~Huang, ``Nonlinear evolution equations as fast
  and exact solvers of estimation problems,'' \emph{IEEE Transactions on Signal
  Processing}, vol.~53, no.~2, pp. 484--498, 2005.

\bibitem{1DVT}
L.~{Condat}, ``A direct algorithm for 1-d total variation denoising,''
  \emph{IEEE Signal Processing Letters}, vol.~20, no.~11, pp. 1054--1057, Nov
  2013.

\bibitem{louchet:hal-00457763}
\BIBentryALTinterwordspacing
C.~Louchet and L.~Moisan, ``{Total Variation as a local filter},'' \emph{{SIAM
  Journal on Imaging Sciences}}, vol.~4, no.~2, pp. 651--694, 2011. [Online].
  Available: \url{https://hal.archives-ouvertes.fr/hal-00457763}
\BIBentrySTDinterwordspacing

\bibitem{chambolle2004algorithm}
A.~Chambolle, ``An algorithm for total variation minimization and
  applications,'' \emph{Journal of Mathematical imaging and vision}, vol.~20,
  no. 1-2, pp. 89--97, 2004.

\bibitem{wen2011parameter}
Y.-W. Wen and R.~H. Chan, ``Parameter selection for total-variation-based image
  restoration using discrepancy principle,'' \emph{IEEE Transactions on Image
  Processing}, vol.~21, no.~4, pp. 1770--1781, 2011.

\bibitem{morozov2012methods}
V.~A. Morozov, \emph{Methods for solving incorrectly posed problems}.\hskip 1em
  plus 0.5em minus 0.4em\relax Springer Science \& Business Media, 2012.

\bibitem{hashemi2015adaptive}
S.~Hashemi, S.~Beheshti, R.~S. Cobbold, and N.~Paul, ``Adaptive updating of
  regularization parameters,'' \emph{Signal Processing}, vol. 113, pp.
  228--233, 2015.

\bibitem{donoho1994ideal}
D.~L. Donoho and J.~M. Johnstone, ``Ideal spatial adaptation by wavelet
  shrinkage,'' \emph{biometrika}, vol.~81, no.~3, pp. 425--455, 1994.

\bibitem{bioucas2006bayesian}
J.~M. Bioucas-Dias, ``Bayesian wavelet-based image deconvolution: a gem
  algorithm exploiting a class of heavy-tailed priors,'' \emph{IEEE
  Transactions on Image Processing}, vol.~15, no.~4, pp. 937--951, 2006.

\bibitem{stein1981estimation}
C.~M. Stein, ``Estimation of the mean of a multivariate normal distribution,''
  \emph{The annals of Statistics}, pp. 1135--1151, 1981.

\bibitem{sardy2016threshold}
S.~Sardy and H.~Monajemi, ``Efficient threshold selection for multivariate
  total variation denoising,'' \emph{Journal of Computational and Graphical
  Statistics}, vol.~28, no.~1, pp. 23--35, 2019.

\bibitem{hansen1992analysis}
P.~C. Hansen, ``Analysis of discrete ill-posed problems by means of the
  l-curve,'' \emph{SIAM review}, vol.~34, no.~4, pp. 561--580, 1992.

\bibitem{kempe2004statistical}
A.~Kempe, ``Statistical analysis of discontinuous phenomena with potts
  functionals,'' Ph.D. dissertation, lmu, 2004.

\bibitem{DonohoDavidL1995AtUS}
\BIBentryALTinterwordspacing
D.~L. Donoho and I.~M. Johnstone, ``\BIBforeignlanguage{eng}{Adapting to
  unknown smoothness via wavelet shrinkage},''
  \emph{\BIBforeignlanguage{eng}{Journal of the American Statistical
  Association}}, vol.~90, no. 432, pp. 1200--1224, 1995. [Online]. Available:
  \url{http://www.tandfonline.com/doi/abs/10.1080/01621459.1995.10476626}
\BIBentrySTDinterwordspacing

\bibitem{arlot2010survey}
S.~Arlot, A.~Celisse \emph{et~al.}, ``A survey of cross-validation procedures
  for model selection,'' \emph{Statistics surveys}, vol.~4, pp. 40--79, 2010.

\end{thebibliography}

\end{document}